\documentclass{LMCS}

\RequirePackage[T1]{fontenc}        
\RequirePackage{xspace}             
\RequirePackage{array}              
\RequirePackage{amssymb}
\RequirePackage{mathpartir}

\newcommand{\N}[1]{\ensuremath{#1}}
\newcommand{\br}{\par\medskip\noindent}

\usepackage{macros}
\usepackage{enumerate,hyperref}

\newcommand{\ignore}[1]{}
\newcommand{\lmcs}[1]{#1}

\def\doi{6 (2:4) 2010}
\lmcsheading%
{\doi}
{1--33}
{}
{}
{Dec.~18, 2009}
{Jun.~23, 2010}
{}

\begin{document}
\title[Tableaux for Simple Type Theory]{Analytic Tableaux for Simple Type Theory\\ and its First-Order Fragment}
\author[C.~E.~Brown]{Chad E.~Brown\rsuper a}
\address{{\lsuper{a,b}}Saarland University}
\email{\{cebrown,smolka\}@ps.uni-saarland.de}
\author[G.~Smolka]{Gert Smolka\rsuper b}
\address{\vskip-6 pt}

\begin{abstract}
  \noindent We study simple type theory with primitive equality
  (STT) and its first-order fragment EFO, which restricts
  equality and quantification to base types but retains
  lambda abstraction and higher-order variables.  As
  deductive system we employ a cut-free tableau calculus.
  We consider completeness, compactness, and existence of
  countable models.
  We prove these properties for STT with respect to Henkin
  models and for EFO with respect to standard models.
  We also show that the tableau system yields a decision
  procedure for three EFO fragments.
\end{abstract}

\maketitle

\section{Introduction}

Church's type theory~\cite{Church40} is a basic
formulation of higher-order logic.
Henkin~\cite{Henkin50} found a natural class of models
for which Church's Hilbert-style proof system turned
out to be complete.  Equality, originally expressed
with higher-order quantification, was later identified
as the primary primitive of the
theory~\cite{Henkin63,Andrews72a,AndrewsBook}.
In this paper we consider simple type theory with
primitive equality but without descriptions or choice.
We call this system STT for simple type
theory.  The semantics of STT is given by Henkin
models with equality.

Modern proof theory started with
Gentzen's~\cite{Gentzen1935} invention of a cut-free
sequent calculus for first-order logic.  While Gentzen
proved a cut-elimination theorem for his calculus,
Smullyan~\cite{SmullyanBook} found an elegant technique
(abstract consistency classes) for proving the
completeness of cut-free first-order calculi.
Smullyan~\cite{SmullyanBook} found it advantageous to
work with a refutation-oriented variant of Gentzen's
sequent calculi~\cite{Gentzen1935} known as tableau
calculi~\cite{Beth1955,Hintikka1955,SmullyanBook}.

The development of complete cut-free proof systems for
simple type theory turned out to be hard.  In 1953,
Takeuti~\cite{Takeuti53} introduced a sequent calculus
for a version of simple type theory without primitive equality
and conjectured that cut elimination holds for
this calculus.  Gentzen's~\cite{Gentzen1935} inductive
proof of cut-elimination for first-order sequent
calculi does not generalize to the higher-order case
since instances of formulas may be more complex than
the formula itself.
Moreover, Henkin's~\cite{Henkin50} completeness proof
cannot be adapted for cut-free systems.  Takeuti's
conjecture was answered positively by
Tait~\cite{Tait66} for second-order logic, by
Takahashi~\cite{Takahashi67} and
Prawitz~\cite{Prawitz68} for higher-order logic without
extensionality, and by Takahashi~\cite{Takahashi68} for
higher-order logic with extensionality.  
Building on
the possible-values technique of
Takahashi~\cite{Takahashi67} and
Prawitz~\cite{Prawitz68}, Takeuti \cite{Takeuti75}
finally proves Henkin completeness of a cut-free
sequent calculus with extensionality.


The first cut-elimination result for a calculus similar
to Church's type theory was obtained by
Andrews~\cite{Andrews71} in 1971.  Andrews considers
elementary type theory (Church's type theory without
equality, extensionality, infinity, and choice) and
proves that a cut-free sequent calculus is complete
relative to a Hilbert-style proof system.  
Andrews' proof employs both the possible-values
technique~\cite{Takahashi67,Prawitz68} and the abstract
consistency technique~\cite{SmullyanBook}.
In 2004 Benzm\"uller, Brown and Kohlhase~\cite{BBKweb04} gave
a completeness proof for an extensional cut-free sequent calculus.
The constructions in~\cite{BBKweb04} also employ abstract consistency
and possible values.

None of the cut-free calculi discussed above has
equality as a primitive.  Following Leibniz, one can
define equality of $a$ and $b$ to hold whenever $a$ and
$b$ satisfy the same properties.  While this yields
equality in standard models (full function spaces),
there are Henkin models where this is not the
case as was shown by Andrews~\cite{Andrews72a}.  
A particularly disturbing fact about the model Andrews constructs
is that while it is extensional (indeed, it is a Henkin model),
it does not satisfy a formula corresponding to extensionality
(formulated using Leibniz equality).
In~\cite{Andrews72a} Andrews gives
a definition of a {\em general model} which is essentially a Henkin model
with equality.
This notion of a general model was generalized to include non-extensional models in~\cite{BBK04}
and a condition called property $\mathfrak{q}$
was explicitly included to ensure Leibniz equality is the same as semantic equality.
The constructions of Prawitz, Takahashi, Andrews and Takeuti described above
do not produce models guaranteed to satisfy property $\mathfrak{q}$.
A similar generalization of Henkin models to non-extensional models is given by Muskens~\cite{Muskens07}
but without a condition like property $\mathfrak{q}$.  Muskens uses the Prawitz-Takahashi method
to prove completeness of a cut-free sequent calculus for a formulation of elementary type theory
via a model existence theorem, again producing a model in which Leibniz equality may not be the same as semantic equality.
The models constructed in~\cite{BBK04} do satisfy property $\mathfrak{q}$,
as do the models constructed in~\cite{BBKweb04}.

In addition
to the model-theoretic complication, defined equality also 
destroys the cut-freeness of a proof system.  As shown
in~\cite{BBK2009} any use of Leibniz equality to say
two terms are equal provides for the simulation of 
cut.\footnote{From a Leibniz formula of the form $\forall p.p s\to p t$ one can easily infer $u\to u$ for any formula $u$, and then use $u$ as a formula introduced by cut.} 
Hence calculi that define equality as Leibniz equality
cannot claim to provide cut-free equational reasoning.
In the context of resolution,
Benzm\"uller gives serious consideration to primitive equality
and its relationship to Leibniz equality in his 1999 doctoral thesis~\cite{Benzmuller99a} (see also~\cite{Benzmuller99b}).
The completeness proofs there are relative to an assumption that corresponds to cut.

The first completeness proof for a cut-free proof system
for extensional simple type theory with primitive equality
relative to Henkin models was given by
Brown in his 2004 doctoral thesis~\cite{Brown2004a}
(later published as a book~\cite{BrownARHO}).  Brown
proves the Henkin completeness of a novel one-sided
sequent calculus with primitive equality.  His model
construction starts with Andrews'~\cite{Andrews71}
non-extensional possible-values relations and then
obtains a structure isomorphic to a Henkin model by
taking a quotient with respect to a partial equivalence
relation.  Finally, abstract consistency
classes~\cite{SmullyanBook,Andrews71} are used to
obtain the completeness result.  The equality-based
decomposition rules of Brown's sequent calculus have
commonalities with the unification rules of the systems
of Kohlhase~\cite{KohlhaseTableaux1995} and
Benzm\"uller~\cite{Benzmuller99b}.  Note, however, that
the completeness proofs of Kohlhase and Benzm\"uller
assume the presence of cut.

In this paper we improve and simplify Brown's
result~\cite{BrownARHO}.  
For the proof system we
switch to a cut-free tableau calculus $\TS$ that employs an
abstract normalization operator.  With the
normalization operator we hide the details of lambda
conversion from the tableau calculus and most of the
completeness proof.  For the completeness proof we use
the new notion of a value system to directly construct
surjective Henkin models.  Value systems are logical
relations~\cite{Statman85a} providing a relational
semantics for simply-typed lambda calculus.  The
inspiration for value systems came from the
possible-values relations used
in~\cite{BrownARHO,BrownSmolkaBasic,BrownSmolkaEFO}.
In contrast to Henkin models, which obtain values for
terms by induction on terms, value systems obtain
values for terms by induction on types.  Induction on
types, which is crucial for our proofs, has the
advantage of hiding the presence of the lambda binder.
As a result, only a single lemma of our completeness
proof deals explicitly with lambda abstractions and
substitutions.

Once we have established the results for STT, we turn
to its first-order fragment EFO (for extended
first-order), which restricts equality and
quantification to base types but retains lambda
abstraction and higher-order variables.  EFO contains
the usual first-order formulas but also contains
formulas that are not first-order in the traditional
sense.  For instance, a formula $p(\lam{x}{\neg fx})$
is EFO even though the predicate $p$ is applied to a
$\lambda$-abstraction and the negation appears embedded
in a nontrivial way.  We sharpen the results for STT by
proving that they hold for EFO with respect to standard
models and for a constrained rule for the universal
quantifier (first published in~\cite{BrownSmolkaEFO}).

Finally, we consider three decidable fragments of EFO:
the lambda-free fragment, the pure fragment
(disequations between simply typed $\lambda$-terms not
involving logic), and the Bernays-Sch\"onfinkel-Ramsey
fragment.  For each of these fragments, decidability
follows from termination of the tableau calculus for EFO
(first published in~\cite{BrownSmolkaBasic}
and~\cite{BrownSmolkaEFO}).

\section{Basic Definitions}

We assume a countable set of \emph{base types}
($\beta$).  \emph{Types} ($\sigma$, $\tau$, $\mu$) are
defined inductively: (1)~every base type is a type;
(2)~if $\sigma$ and $\tau$ are types, then $\sigma\tau$
is a type.  We assume a countable set of \emph{names}
($x$, $y$), where every name comes with a unique type,
and where for every type there are infinitely many names of this 
type.\footnote{Later we will partition names into variables and logical constants.} 
\emph{Terms} ($s$, $t$, $u$, $v$)
are defined inductively: (1)~every name is a term;
(2)~if $s$ is a term of type $\tau\mu$ and $t$ is a
term of type $\tau$, then $st$ is a term of type $\mu$;
(3)~if $x$ is a name of type $\sigma$ and $t$ is a term
of type $\tau$, then $\lam{x}t$ is a term of type
$\sigma\tau$.  We write \emph{$s:\sigma$} to say that
$s$ is a term of type $\sigma$.  Moreover, we write
\emph{$\Wff_\sigma$} for the set of all terms of type
$\sigma$.  
We assume that the set of types and the set of terms
are disjoint.

A \emph{frame} is a function $\mcd$ that maps every
type to a nonempty set such that $\mcd(\sigma\tau)$ is
a set of total functions from $\mcd\sigma$ to
$\mcd\tau$ for all types $\sigma$, $\tau$ (i.e.,
$\mcd(\sigma\tau)\incl(\mcd\sigma\to\mcd\tau)$).
An
\emph{assignment} into a frame $\mcd$ is a function
$\mci$ that extends $\mcd$ (i.e., $\mcd\incl\mci$) and
maps every name $x:\sigma$ to an element of $\mcd\sigma$
(i.e., $\mci x\in\mcd\sigma$).  If~$\mci$ is an
assignment into a frame $\mcd$, $x:\sigma$ is a
name, and $a\in\mcd\sigma$,
then~\emph{$\subst\mci{x}a$} denotes the assignment
into $\mcd$ that agrees everywhere with~$\mci$ but
possibly on~$x$ where it yields $a$.  For every frame
$\mcd$ we define a function \emph{$\hat{~}$} that for
every assignment~$\mci$ into $\mcd$ yields a
function $\hat\mci$ that for some terms $s:\sigma$
returns an element of $\mcd\sigma$.  The definition is
by induction on terms.
\begin{align*}
  \hat\mci x&\eqdef\mci x \\
  \hat\mci(st)&\eqdef fa
  &&\text{if \ $\hat\mci s=f$ \ and \ $\hat\mci t=a$}  \\
  \hat\mci(\lam{x}s)&\eqdef f
  &&\text{if \ $\lam{x}s:\sigma\tau$, \ 
    $f\in\mcd(\sigma\tau)$, \ and \ 
    $\forall a\in\mcd\sigma\col~~
    \widehat{\subst\mci{x}a}s=fa$}  
\end{align*}
We call $\hat\mci$ the \emph{evaluation function} of
$\mci$.  
The evaluation function may be partial since in the last clause  
of the definition even assuming there is some function 
$f$ such that $\widehat{\subst\mci{x}a}s=fa$ for every $a\in\mcd\sigma$, 
this $f$ may not be in $\mcd(\sigma\tau)$.  In such a case, $\hat\mci$ 
will not be defined on $\lam{x}s$.  
Of course, in such a case $\hat\mci$ will also not be defined on a term of the form $(\lam{x}s)t$ since
the second clause of the definition will fail. 
An \emph{interpretation} is an
assignment whose evaluation function is defined on
all terms.  An assignment $\mci$ is
\emph{surjective} if for every type $\sigma$ and every
value $a\in\mci\sigma$ there exists a term $s:\sigma$
such that $\hat\mci s=a$.

\begin{prop}
  Let $\mci$ be an interpretation, $x:\sigma$,
  and $a\in\mci\sigma$.  Then~$\subst\mci{x}a$ is an
  interpretation.
\end{prop}


\begin{prop}
  If $\mci$ is a surjective interpretation, then
  $\mci\sigma$ is a countable set for every type
  $\sigma$.
\end{prop}

A \emph{standard frame} is a frame $\mcd$ such that
$\mcd(\sigma\tau)=(\mcd\sigma\to\mcd\tau)$ for all
types $\sigma$, $\tau$.  A \emph{standard
  interpretation} is an assignment into a standard
frame.  Note that every standard interpretation is, in fact, an
interpretation.

We assume a \emph{normalization operator $\nf{\cdot}$} that 
provides for lambda conversion.  The normalization
operator $\nf{\cdot}$ must be a type preserving total 
function from terms to terms.  We call $\nf{s}$ the
\emph{normal form of $s$} and say that $s$ is
\emph{normal} if $\nf{s}=s$.  One possible
normalization operator is a function that for every
term $s$ return a $\beta$-normal term that can be
obtained from $s$ by $\beta$-reduction.  We will not
commit to a particular normalization operator but state
explicitly the properties we require for our results.
To start, we require the following properties:
\begin{description}
\item[{N1}~] $\nf{\nf{s}}=\nf{s}$
\item[{N2}~] $[[s]t]=\nf{st}$
\item[{N3}~] $\nf{x\ddd s n}=x\nf{s_1}\dots\nf{s_n}$ 
  \quad if $x\ddd s n:\beta$ and $n\ge0$
\item[{N4}~] $\hat\mci\nf{s}=\hat\mci{s}$
  \quad if $\mci$ is an interpretation
\end{description}

\begin{prop}
  $x\ddd{s}n:\beta$ is normal iff $\dd{s}n$ are normal.
\end{prop}


For the proofs of Lemma~\ref{lem-admissibility} and
Theorem~\ref{theo-admissible-interpretations} we need
further properties of the normalization operator that
can only be expressed with substitutions.  A
\emph{substitution} is a type preserving partial
function from names to terms.  If $\theta$ is a
substitution, $x$ is a name, and $s$ is a term that has
the same type as $x$, we write \emph{$\subst\theta x
  s$} for the substitution that agrees everywhere
with~$\theta$ but possibly on $x$ where it yields $s$.
We assume that every substitution $\theta$ can be
extended to a type preserving total function
\emph{$\hat\theta$} from terms to terms such that the
following conditions hold:
\enlargethispage*{5mm} 
\begin{description}
\item[{S1}~] $\hat\theta x=\Cond{x\in\Dom\theta}{\theta{x}}{x}$
\item[{S2}~] $\hat\theta(st)=(\hat\theta{s})(\hat\theta{t})$
\item[{S3}~] $[(\hat\theta(\lam{x}s){})t]=[\widehat{\subst\theta{x}t}s]$
\item[{S4}~] $\nf{\hat\eset s}=\nf{s}$
\end{description}
Note that $\eset$ (the empty set) is the substitution
that is undefined on every name.

\section{Value Systems}
\label{sec:value-sys}

We introduce value systems as a tool for constructing
surjective interpretations.  Value systems are
logical relations inspired by the possible-values
relations used
in~\cite{BrownARHO,BrownSmolkaEFO,BrownSmolkaBasic}.

A \emph{value system} is a function $\canbe$ that maps
every base type $\beta$ to a binary
relation~$\canbe_\beta$ such that
$\Dom(\canbe_\beta)\incl\Wff_\beta$ and $s\canbe_\beta
a$ iff $\nf{s}\canbe_\beta a$.  For every value
system~$\canbe$ we define by induction on types:
\begin{align*}
  \N{\mcd\sigma}&\eqdef\Ran(\canbe_\sigma)\\
  \N{\canbe_{\sigma\tau}}&\eqdef\mset{(s,f)\in\Wff_{\sigma\tau}\times(\mcd\sigma\to\mcd\tau)}
  {\forall(t,a)\in\canbe_\sigma\col~(st,fa)\in\canbe_\tau}
\end{align*}
Note that
$\mcd(\sigma\tau)\incl(\mcd\sigma\to\mcd\tau)$ for all
types $\sigma\tau$.  We usually drop the type index in
$s\canbe_\sigma a$ and read $s\canbe a$ as $s$ can be
$a$ or $a$ is a \emph{possible value} for $s$.

\begin{prop}
  \label{prop-norm-poss-value}
  For every value system: \
  $s\canbe_\sigma a$ iff $\nf{s}\canbe_\sigma a$.
\end{prop}

\begin{proof}
  By induction on $\sigma$.  For base types the claim
  holds by the definition of value systems.  Let
  $\sigma=\tau\mu$.
  For all $s\in\Wff_\sigma$, $t\in\Wff_\tau$, $a\in\mcd\tau\to\mcd\mu$, and $b\in\mcd\tau$,
  $$st\canbe_\mu ab
  {\mbox{ iff }}
  \nf{st}\canbe_\mu ab
  {\mbox{ iff }}
  \nf{\nf{s}t}\canbe_\mu ab
  {\mbox{ iff }}
  \nf{s}t\canbe_\mu ab$$
  by the inductive hypothesis and N2.
  Hence $s\canbe_\sigma a$ iff $\nf{s}\canbe a$.
%
\end{proof}

A value system $\canbe$ is \emph{functional} if
$\canbe_\beta$ is a functional relation for every base
type $\beta$.
(That is, for each $s\in\Wff_\beta$ there is at most one $b$ such that $s\canbe b$.) 

\begin{prop}
  \label{prop-functional-vs}
  If $\canbe$ is functional, then $\canbe_\sigma$ is a
  functional relation for every type~$\sigma$.
\end{prop}

\begin{proof}
  By induction on $\sigma$.  For $\sigma=\beta$, the
  claim is trivial.  Let $\sigma=\tau\mu$ and
  $s\canbe_{\tau\mu}f,g$.  We show $f=g$.  Let
  $a\in\mcd\tau$.  Then $t\canbe_\tau a$ for some $t$.
  Now $st\canbe_\mu fa,ga$.  By inductive hypothesis
  $fa=ga$.
\end{proof}

A value system $\canbe$ is \emph{total} if
$x\in\Dom\canbe_\sigma$ for every name $x:\sigma$.  An
assignment $\mci$ is \emph{admissible} for a value
system $\canbe$ if $\mci\sigma=\mcd\sigma$ for all
types $\sigma$ and $x\canbe\mci x$ for all names $x$.
(Recall that $\canbe$ is used to define $\mcd$.) 
Note that every total value system has admissible
assignments.  We will show that admissible
assignments are interpretations that
evaluate terms to possible values.

\begin{lem}
  \label{lem-admissibility}
  Let $\mci$ be an assignment that is admissible
  for a value system $\canbe$ and $\theta$ be a
  substitution such that $\theta{x}\canbe\mci{x}$ for
  all $x\in\Dom\theta$.  Then $s\in\Dom\hat\mci$ and
  $\hat\theta{s}\canbe\hat\mci{s}$ for every term $s$.
\end{lem}

\begin{proof}
  By induction on $s$.  Let $s$ be a term.  Case
  analysis.

  \br $s=x$.  The claim holds by assumption and S1.

  \br $s=tu$.  Then $t\in\Dom\hat\mci$, \
  $\hat\theta{t}\canbe\hat\mci{t}$, \
  $u\in\Dom\hat\mci$, \ and
  $\hat\theta{u}\canbe\hat\mci{u}$ by inductive
  hypothesis.  Thus $s\in\Dom\hat\mci$ and
  $\hat\theta{s}= (\hat\theta{t})(\hat\theta{u})\canbe
  (\hat\mci{t})(\hat\mci{u})=\hat\mci{s}$ using S2.

  \br $s=\lam{x}t$, $x:\sigma$ and $t:\tau$.  
  We need to prove $s\in\Dom\hat\mci$ and $\hat\theta{s}\canbe\hat\mci{s}$.
  First we prove
  \begin{equation}\label{lem-adm-funcase}
    t\in\Dom\widehat{\subst\mci{x}{a}} {\mbox{ and }} (\hat\theta{s})u\canbe \widehat{\subst\mci{x}a}t {\mbox{ whenever }} u\canbe_\sigma a.  
  \end{equation}
  Let $u\canbe_\sigma a$.  
  By inductive hypothesis we have $t\in\Dom\widehat{\subst\mci{x}a}$
  and $\widehat{\subst\theta{x}u}t\canbe
  \widehat{\subst\mci{x}a}t$.
  Now
  $\nf{(\hat\theta{s})u}=
  \nf{\widehat{\subst\theta{x}u}t}\canbe
  \widehat{\subst\mci{x}a}t$ using~S3.
  Using Proposition~\ref{prop-norm-poss-value} we conclude (\ref{lem-adm-funcase}) holds.

  By definition of $\mcd\sigma$ for every $a\in\mcd\sigma$ there is a $u$ such that $u\canbe a$.
  Using this and (\ref{lem-adm-funcase}) we know $t\in\Dom\widehat{\subst\mci{x}a}$ for every $a\in\mcd\sigma$.
  Let $f:\mcd\sigma\to\mcd\tau$ be defined by $fa = \widehat{\subst\mci{x}{a}}{t}$
  for each $a\in\mci\sigma$.  
  For all $u\canbe_\sigma a$ we have $(\hat\theta{s})u\canbe fa$ by (\ref{lem-adm-funcase}).
  Hence $\hat\theta{s}\canbe f$.
  This implies $f\in\mcd{(\sigma\tau)}$, $s\in\Dom\hat\mci$, $\hat\mci s = f$
  and $\hat\theta{s}\canbe \hat\mci s$ as desired.
\end{proof}

\begin{thm}
  \label{theo-admissible-interpretations}
  Let $\mci$ be an assignment that is admissible
  for a value system~$\canbe$.  Then $\mci$ is an
  interpretation such that $s\canbe\hat\mci s$ for all
  terms $s$.  Furthermore, $\mci$~is surjective
  if~$\canbe$ is functional.
\end{thm}

\begin{proof}
  Follows from Lemma~\ref{lem-admissibility} with
  Proposition~\ref{prop-norm-poss-value} and S4.
  To prove the second claim, let $a\in\mcd\sigma$ be given.  
  By definition of $\mcd$ there is some $s$ such that $s\canbe a$.
  Since $s\canbe\hat\mci s$ we know $\hat\mci s = a$ by
  Proposition~\ref{prop-functional-vs}. 
\end{proof}





\section{Simple Type Theory}

We now define the terms and semantics of simple type theory (\emph{STT}).
We fix a base type $\N{o}$ for the truth values and a
name $\N{\neg}:oo$ for negation.
Moreover, we fix for every type
$\sigma$ a name $\N{=_\sigma}:\sigma\sigma o$ for the
identity predicate for $\sigma$.
An assignment
$\mci$ is \emph{logical} if $\mci o=\set{0,1}$,
$\mci(\neg)$ is the negation
function and $\mci(=_\sigma)$ is the identity
predicate for $\sigma$.
We refer to the base types
different from $o$ as \emph{sorts}, to the names
$\neg$ and $=_\sigma$ as \emph{logical constants}, 
and to all other names as \emph{variables}.  
From now on \emph{$x$} will range
over variables. 
Moreover, \emph{$c$} will range over logical constants
and \emph{$\alpha$} will range over sorts.

A \emph{formula} is a term of type $o$.  We employ
infix notation for formulas obtained with 
$=_\sigma$ and often write \emph{equations} $s=_\sigma
t$ without the type index.  
We write \emph{$s\neq t$} for $\neg (s{=}t)$
and speak of a \emph{disequation}.
Note that
quantified formulas $\forall x.s$ can be expressed as
equations $(\lam{x}s)=(\lam{x}x=x)$.

A logical interpretation $\mci$ \emph{satisfies}
a formula $s$ if $\hat\mci s=1$.  A \emph{model} of a
set of formulas $A$ is a logical interpretation
that satisfies every formula~$s\in A$.  A set of
formulas is \emph{satisfiable} if it has a model.

\section{Tableau Calculus}

We now give a deductive calculus for STT.
A \emph{branch} is a set of normal formulas.  The
\emph{tableau calculus $\TS$} operates on finite branches
and employs the rules shown in
Figure~\ref{fig-tableau-rules}.
\begin{figure}[t]
\begin{mathpar}
  \inferrule*[left=\emph{\TRDN}~]{\neg\neg s}{s}
  \and
  \inferrule*[left=\emph{\TRBQ}~]{s =_ot}{s\,,\,t~\mid~\neg s\,,\,\neg t}
  \and
  \inferrule*[left=\emph{\TRBE}~]{s\neq_ot}{s\,,\,\neg t~\mid~\neg s\,,\,t}
  \\
  \inferrule*[left=\emph{\TRFQ}~,right=~$u:\sigma$ normal]
  {s =_{\sigma\tau} t}{\nf{su} =\nf{tu}}
  \and
  \inferrule*[left=\emph{\TRFE}~,right=~$x:\sigma$ fresh]
  {s\neq_{\sigma\tau} t}{\nf{sx}\neq\nf{tx}}
  \\
  \inferrule*[left=\emph{\TRMat}~,right=~$n\geq 0$] {xs_1\dots
    s_n\,,\,\neg xt_1\dots t_n} {s_1\neq t_1\mid\dots\mid s_n\neq t_n}
  \and
  \inferrule*[left=\emph{\TRDec}~,right=~$n\geq 0$] {xs_1\dots
    s_n\neq_\alpha xt_1\dots t_n} {s_1\neq t_1\mid\dots\mid s_n\neq t_n}
  \\\
  \inferrule*[left=\emph{\TRCon}~]
  {s=_\alpha t\,,\,u\neq_\alpha v}
  {s\neq u\,,\,t\neq u\mid s\neq v\,,\,t\neq v}
\end{mathpar}
\caption{Tableau rules for STT}
\label{fig-tableau-rules}
\end{figure}
The side condition ``$x$~fresh'' of rule \TRFE requires
that $x$ does not occur free in the branch the rule is
applied to.  
We say a branch $A$ is \emph{closed} if $x,\neg x\in A$ for some variable $x:o$
or if $x\not=_\iota x\in A$ for some variable $x:\iota$.
Note that $A$ is closed if and only if either the $\TRMat$ or $\TRDec$ rule 
applies with $n=0$.
We impose the following restrictions:
\begin{enumerate}[(1)]
\item We only admit rule instances $A/\ddd A n$ where $A$ is not closed.
\item \TRFE can only be applied to a disequation
  $(s{\neq} t)\in A$ if there is no variable $x$ such that 
  $(\nf{sx}\neq\nf{tx})\in A$.
\end{enumerate}
The set of \emph{refutable branches} is
defined inductively: if $A/\ddd A n$ is
an instance of a rule of~$\TS$ and $\dd A n$ are
refutable, then $A$ is refutable.
Note that the base cases of this inductive definition are when $n=0$.
The rules where $n$ may be $0$ are $\TRMat$ and $\TRDec$.
Figure~\ref{fig:refutation} shows a refutation
in~$\TS$.

\begin{figure}[t]
\begin{equation*}
  \begin{array}{c}
    pf,~\neg p(\lam{x}{\neg\neg fx}) \\
    {\mbox{[${\emph{\TRMat}}$]}} \\
    f\neq(\lam{x}{\neg\neg fx}) \\
    {\mbox{[${\emph{\TRFE}}$]}} \\
    fx\neq\neg\neg fx \\
    {\mbox{[${\emph{\TRBE}}$]}} \\
    \hline
    \begin{array}{c|c}
      \begin{array}{c}
        fx,~\neg\neg\neg fx\\
        {\mbox{[${\emph{\TRDN}}$]}} \\
        \neg fx\\
        {\mbox{[${\emph{\TRMat}}$]}} \\
        x\neq x
        \\ {\mbox{[${\emph{\TRDec}}$]}}
      \end{array}
      &
      \begin{array}{c}
        \neg fx,~\neg\neg fx \\
        {\mbox{[${\emph{\TRDN}}$]}} \\
        fx\\
        {\mbox{[${\emph{\TRMat}}$]}} \\
        x\neq x
        \\ {\mbox{[${\emph{\TRDec}}$]}}
      \end{array}
    \end{array}
  \end{array}
\end{equation*}  
\caption{Tableau refuting $\{pf,\neg p(\lam{x}{\neg\neg fx})\}$
  where $p:(\alpha o)o$ and $f:\alpha o$}
\label{fig:refutation}
\end{figure}

A remark on the names of the rules: \TRMat is called
the mating rule, \TRDec the decomposition rule, \TRCon
the confrontation rule, \TRBQ the Boolean equality rule, \TRBE
the Boolean extensionality rule, \TRFQ the functional equality
rule, and \TRFE the functional extensionality rule.

\begin{prop}[Soundness] \label{prop:ts-sound}
  Every refutable branch is unsatisfiable.
\end{prop}

\begin{proof}
  Let $A/\ddd A n$ be an instance of a rule of $\TS$
  such that $A$ is satisfiable.  It suffices to show
  that one of the branches $\dd A n$ is satisfiable.
  Straightforward.
\end{proof}

We will show that the tableau calculus $\TS$ is
\emph{complete}, that is, can refute every finite
unsatisfiable branch.  The rules of $\TS$ are designed
such that we obtain a strong completeness result.
For practical purposes one can of course include rules 
that close branches including $s,\neg s$
or $s\neq s$.

To avoid redundancy, our definition of STT only covers the logical
constants $\neg$ and $=_\sigma$.
Adding further constants such as $\land$, $\lor$, $\to$,
$\forall_{\!\sigma}$ and $\exists_\sigma$ is straightforward.
In fact, all logical constants can be expressed with the identities
$=_\sigma$~\cite{AndrewsBook}.  We have included $\neg$ since we need
it for the formulation of the tableau calculus.  The refutation
in Figure~\ref{fig:refutationneg}
suggests that the elimination of $\neg$ is not straightforward.

\begin{figure}[t]
\begin{equation*}
  \begin{array}{c}
    (\lam{x}{x}) = \lam{x}{y} \\
    {\mbox{[${\emph{\TRFQ}}$ with $x$]}} \\
    x =_o y \\
    {\mbox{[${\emph{\TRBQ}}$]}} \\
    \hline
    \begin{array}{c|c}
      \begin{array}{c}
        x,y\\
        {\mbox{[${\emph{\TRFQ}}$ with $\neg x$]}} \\
    (\neg x) =_o y \\
    {\mbox{[${\emph{\TRBQ}}$]}} \\
        \hline
        \begin{array}{c|c}
          \begin{array}{c}
            \neg x,y
            \\ {\mbox{[${\emph{\TRMat}}$]}}
          \end{array}
          &
          \begin{array}{c}
            \neg \neg x,\neg y
            \\ {\mbox{[${\emph{\TRMat}}$]}}
          \end{array}
        \end{array}
      \end{array}          
      &
      \begin{array}{c}
        \neg x,\neg y\\
        {\mbox{[${\emph{\TRFQ}}$ with $\neg x$]}} \\
    (\neg x) =_o y \\
    {\mbox{[${\emph{\TRBQ}}$]}} \\
        \hline
        \begin{array}{c|c}
          \begin{array}{c}
            \neg x,y
            \\ {\mbox{[${\emph{\TRMat}}$]}}
          \end{array}
          &
          \begin{array}{c}
            \neg \neg x,\neg y\\
            {\mbox{[${\emph{\TRDN}}$]}} \\
            x
            \\ {\mbox{[${\emph{\TRMat}}$]}}
          \end{array}
        \end{array}
      \end{array}          
    \end{array}
  \end{array}
\end{equation*}  
\caption{Tableau refuting $(\lam{x}{x}) = \lam{x}{y}$ where $x,y:o$}
\label{fig:refutationneg}
\end{figure}

\section{Evidence}

A branch $E$ is \emph{evident} if it satisfies the
\emph{evidence conditions} in
Figure~\ref{fig:evidence}.  The evidence conditions
correspond to the tableau rules and are designed such
that every branch that is closed under the
tableau rules is either closed or evident.  We will show that evident
branches are satisfiable.

\begin{figure}[t]
  \renewcommand{\arraystretch}{1.4}
  \begin{tabular}{c>{\raggedright}p{120mm}}
    \emph{\EDN}&If $\neg\neg s$ is in $E$, then $s$ is in $E$.
    \tabularnewline
    \emph{\EBQ}&If $s =_o t$ is in $E$, 
    then either $s$ and $t$ are in $E$ or $\neg s$ and $\neg t$ are in $E$.
    \tabularnewline
    \emph{\EBE}&If $s\neq_o t$ is in $E$, 
    then either $s$ and $\neg t$ are in $E$ or $\neg s$ and $t$ are in $E$.
    \tabularnewline
    \emph{\EFQ}&If $s =_{\sigma\tau} t$ is in $E$, 
    then $\nf{su}=\nf{tu}$ is in $E$ for every normal $u:\sigma$.
    \tabularnewline
    \emph{\EFE}&If $s\neq_{\sigma\tau} t$ is in $E$, 
    then $\nf{sx}\neq\nf{tx}$ is in $E$ for some variable $x$.
    \tabularnewline
    \emph{\EMat}&If $x\ddd s n$ and $\neg x\ddd t n$ are in $E$,\ignore{\\}
    then $n\ge1$ and $s_i\neq t_i$ is in $E$ for some $i\in\set{1\cld n}$.
    Note that if $n=0$, this means if $\neg x\in E$, then $x\notin E$.
    \tabularnewline
    \emph{\EDec}&If $x\ddd s n\neq_\alpha x\ddd t n$ is in $E$,\ignore{\\}
    then $n\ge1$ and $s_i\neq t_i$ is in $E$ for some $i\in\set{1\cld n}$.
    Note that if $n=0$, this means $x\neq_\alpha x\notin E$.
    \tabularnewline
    \emph{\ECon}&If $s=_\alpha t$ and $u \neq_\alpha v$  are in $E$,\\
    then either $s\neq u$ and $t\neq u$ are in $E$ 
    or $s\neq v$ and $t\neq v$ are in $E$.
  \end{tabular}
  \caption{Evidence conditions}
  \label{fig:evidence}
\end{figure}

A branch $E$ is \emph{complete} if for every normal
formula $s$ either $s$ or $\neg s$ is in $E$.  The
cut-freeness of $\TS$ shows in the fact that there are
many evident sets that are not complete.  For instance,
$\set{pf,~\neg p(\lam{x}{\neg fx}),~f\neq\lam{x}{\neg
    fx},~ fx\neq\neg fx,~\neg fx}$ is an incomplete
evident branch if $p:(\sigma o)o$.

\subsection{Discriminants}

Given an evident branch $E$, we will construct a value
system whose admissible logical interpretations are
models of $E$.  We start by defining the values for the
sorts, which we call discriminants.
Discriminants first appeared in~\cite{BrownSmolkaBasic}.

Let $E$ be a fixed evident branch in the following. 
A term $u\in\Wff_\alpha$ is \emph{$\alpha$-discriminating in $E$}
if there
is some term $t$ such that either $u\neq_\alpha t$ or
$t\neq_\alpha u$ is in $E$.
An \emph{$\alpha$-discriminant} is a maximal set $a$ of
discriminating terms of type $\alpha$ such that there
is no disequation $s{\neq}t\in E$ such that $s,t\in a$.
We write \emph{$s\notq t$} if $E$ contains the
disequation $s{\neq }t$ or~$t{\neq}s$.

In~\cite{Brown2004a} a sort was interpreted using
maximally compatible sets of terms of the sort (where $s$ and $t$
are compatible unless $s\notq t$).
The idea is that the set $E$ insists that certain terms cannot be
equal, but leaves open that other terms ultimately may be identified
by the interpretation.  In particular, two compatible terms $s$ and $t$ may be identified by
taking a maximally compatible set of terms containing both $s$ and $t$ as a value.
It is not difficult to see that a maximally compatible set is simply the union of an $\alpha$-discriminant
with all terms of sort $\alpha$ that are not $\alpha$-discriminating.
We now find that it is clearer to use $\alpha$-discriminants as values instead of maximally compatible sets.
In particular, it is easier to count the number of $\alpha$-discriminants, as we now show.

\begin{exa}
  Suppose $E=\set{x{\neq}y,\,x{\neq}z,\,y{\neq}z}$ and
  $x,y,z:\alpha$.  There are~3
  \text{$\alpha$-discriminants}: $\set{x}$, $\set{y}$,
  $\set{z}$.
\end{exa}

\begin{exa}
  Suppose $E=\mset{a_n\neq_\alpha b_n}{n\in\NN}$ where
  the $a_n$ and $b_n$ are pairwise distinct variables.
  Then $E$ is evident and there are uncountably many
  \text{$\alpha$-discriminants}.
\end{exa}

\begin{prop}
  \label{prop-finite-discs}
  If $E$ contains exactly $n$ disequations at $\alpha$,
  then there are at most~$2^n$ $\alpha$-discriminants.
  If $E$ contains no disequation at $\alpha$, then
  $\eset$ is the only $\alpha$-discriminant.
\end{prop}

\begin{prop}
  \label{prop-diff-discs}
  Let $a$ and $b$ be different discriminants.
  Then:
  \begin{enumerate}[\em(1)]
  \item $a$ and $b$ are separated by a disequation
    in $E$, that is, there exist terms $s\in a$ and
    $t\in b$ such that $s\notq t$.
  \item $a$ and $b$ are not connected by an
    equation in $E$, that is, there exist no terms
    $s\in a$ and $t\in b$ such that $(s{=}
    t)\in E$.
  \end{enumerate}
\end{prop}

\begin{proof}
  The first claim follows by contradiction.  Suppose
  there are no terms ${s\in a}$ and $t\in b$ such that
  $s\notq t$.  Let $s\in a$.  Then $s\in b$ since $b$
  is a maximal set of discriminating terms.  Thus
  $a\incl b$ and hence $a=b$ since $a$ is maximal.
  Contradiction.

  The second claim also follows by contradiction.
  Suppose there is an equation $(s_1{=} s_2)\in E$ such
  that $s_1\in a$ and $s_2\in b$.  By the first
  claim we have terms $s\in a$ and $t\in b$
  such that $s\notq t$.  By \ECon we have $s_1\notq
  s$ or $s_2\notq t$.  Contradiction since $a$
  and~$b$ are discriminants.
\end{proof}

\subsection{Compatibility}\label{sec:compat}

For our proofs we need an auxiliary notion for evident
branches that we call compatibility.  Let $E$ be a
fixed evident branch in the following.  We define
relations
$\N{\comp_\sigma}\incl\Wff_\sigma\times\Wff_\sigma$ by
induction on types:
\begin{align*}
  \N{s\comp_o t}&\iffdef 
  \set{\nf{s},\neg\nf{t}}\not\subseteq E~\mathrm{and}~
  \set{\neg\nf{s},\nf{t}}\not\subseteq E\\
  \N{s\compa t}&\iffdef
  \mathrm{not}~\nf{s}\notq\nf{t}\\
  \N{s\comp_{\sigma\tau}t}&\iffdef 
  s u\comp_\tau t v~\text{whenever}~u\comp_\sigma v
\end{align*}
We say that $s$ and $t$ are \emph{compatible} if
$s\comp t$.

\begin{lem}[Compatibility]
  \label{lem-compatibility}~\\
  For $n\ge0$ and all terms $s$, $t$, $xs_1\dots s_n$,
  $xt_1\dots t_n$ of type~$\sigma$:
  \begin{enumerate}[\em(1)]
  \item We do not have both $s\comp_\sigma t$ and $\nf{s}\notq\nf{t}$.
  \item Either $xs_1\dots s_n\comp_\sigma xt_1\dots t_n$
    or $\nf{s_i}\notq\nf{t_i}$ for some
    $i\in\set{1\cld n}$.
  \end{enumerate}
\end{lem}

\begin{proof}
  By induction on $\sigma$.  Case analysis.

  $\sigma=o$.  Claim~(1) follows with \EBE.
  Claim~(2) follows with N3 and \EMat.

  $\sigma=\alpha$.  Claim~(1) is trivial.  Claim~(2)
  follows with N3 and \EDec.

  $\sigma=\tau\mu$.  We show~(1) by contradiction.
  Suppose $s\comp_\sigma t$ and $\nf{s}\notq\nf{t}$.
  By \EFE $\nf{\nf{s}x}\notq\nf{\nf{t}x}$ for some
  variable $x$.  By inductive hypothesis~(2) we have
  $x\comp_\tau x$.  Hence $sx\comp_\mu tx$.
  Contradiction by inductive hypothesis~(1) and N2.

  To show~(2), suppose $xs_1\dots s_n\ncomp_\sigma
  xt_1\dots t_n$.  Then there exist terms such that
  $u\comp_\tau v$ and $xs_1\dots s_nu\ncomp_\mu
  xt_1\dots t_nv$.  By inductive hypothesis~(1) we know
  that $\nf{u}\notq\nf{v}$ does not hold.  Hence
  $\nf{s_i}\notq\nf{t_i}$ for some $i\in\set{1\cld n}$
  by inductive hypothesis~(2).
\end{proof}

\section{Model Existence}
\label{sec:model-existence}

Let $E$ be a fixed evident branch.  We define a value
system~$\canbe$ for $E$:
\begin{align*}
  \N{s \canbe_o 0}&\iffdef s\in\Wff_o \text{ and } \nf{s}\notin E\\
  \N{s \canbe_o 1}&\iffdef s\in\Wff_o \text{ and } \neg\nf{s}\notin E\\
  \N{s\canbe_\alpha a}\!&\iffdef 
  s\in\Wff_\alpha,~a \text{ is an $\alpha$-discriminant, and } 
  \nf{s}\in a \text{ if } \nf{s} \text{ is discriminating}
\end{align*}
Note that N1 ensures the property $s\canbe_\beta a$ iff
$\nf{s}\canbe_\beta a$.

\begin{prop}
  \label{prop-mod-ex-o}
  For all variables $x_o$, either $x\canbe 0$ and $\neg x\canbe 1$
  or $x\canbe 1$ and $\neg x\canbe 0$.
  In particular, $\mcd o=\set{0,1}$.
\end{prop}

\begin{proof}
  By $\EMat$ either $x\notin E$ or $\neg x\notin E$.
  If $x\notin E$, then $x\canbe 0$ and $\neg x\canbe 1$ by N3 and $\EDN$.
  If $\neg x\notin E$, then $x\canbe 1$ and $\neg x\canbe 0$ by N3.
\end{proof}

\begin{lem}
  \label{lemma-adm-model}
  A logical assignment is a model of $E$ if it is
  admissible for $\canbe$.
\end{lem}

\begin{proof}
  Let $\mci$ be a logical assignment that is
  admissible for $\canbe$, and let $s\in E$.  By
  Theorem~\ref{theo-admissible-interpretations} we know
  that $\mci$ is an interpretation and that
  $s\canbe_o\hat\mci s$.  Thus $\hat\mci s\neq0$ since
  $s\in E$.  Hence $\hat\mci s=1$.
\end{proof}

It remains to show that $\canbe$ admits logical
interpretations.  First we show that all sets
$\mcd\sigma$ are nonempty.  To do so, we prove that
compatible equi-typed terms have a common value.  A set
$T$ of equi-typed terms is \emph{compatible} if $s\comp
t$ for all terms $s,t\in T$.  We write
\emph{$T\canbe_\sigma a$} if $T\incl\Wff_\sigma$,
$a\in\mcd\sigma$, and $t\canbe a$ for every~$t\in T$.

\begin{lem}[Common Value]
  \label{lem-common-value}
  Let $T\incl\Wff_\sigma$.  Then $T$ is compatible if
  and only if there exists a value~$a$ such that
  $T\canbe_\sigma a$.
\end{lem}

\begin{proof}
  By induction on $\sigma$.  

  \br $\sigma=\alpha,~{\Rightarrow}$.  Let $T$ be
  compatible.  Then there exists an
  $\alpha$-discriminant $a$ that contains all the
  $\alpha$-discriminating terms in $\mset{\nf{t}}{t\in T}$.  
  Clearly, $T\canbe a$. 
  
  \br $\sigma=\alpha,~{\Leftarrow}$. 
  Suppose $T\canbe a$ and $T$ is not compatible.  Then there
  are terms $s,t\in T$ such that
  $(\nf{s}{\neq}\nf{t})\in E$.  Thus $\nf{s}$ and
  $\nf{t}$ cannot be both in $a$.  This contradicts
  $s,t\in T\canbe a$ since $\nf{s}$ and $\nf{t}$ are
  discriminating.

  \br $\sigma=o,~{\Rightarrow}$.  By contraposition.
  Suppose $T\ncanbe0$ and $T\ncanbe1$. Then there are
  terms $s,t\in T$ such that $\nf{s},\neg\nf{t}\in E$.
  Thus $s\ncomp t$.  Hence $T$ is not compatible.

  \br $\sigma=o,~{\Leftarrow}$.  By contraposition.
  Suppose $s\ncomp_o t$ for $s,t\in T$.  Then
  $\nf{s},\neg\nf{t}\in E$ without loss of generality.
  Hence $s\ncanbe0$ and $t\ncanbe1$.  Thus $T\ncanbe0$
  and $T\ncanbe1$.

  \br $\sigma=\tau\mu,~{\Rightarrow}$.  Let $T$ be
  compatible.  We define $T_a:=\mset{ts}{t\in
    T,~s\canbe_\tau a}$ for every value $a\in\mci\tau$
  and show that $T_a$ is compatible.  Let $t_1,t_2\in
  T$ and $s_1,s_2\canbe_\tau a$.  It suffices to show
  $t_1s_1\comp t_2s_2$.  By the inductive hypothesis
  $s_1\comp_\tau s_2$.
  Since $T$ is compatible, $t_1\comp t_2$.
  Hence $t_1s_1\comp t_2s_2$.

  By the inductive hypothesis we now know that for
  every $a\in\mci\tau$ there is a $b\in\mci\mu$ such
  that $T_a\canbe_\mu b$.  Hence there is a function
  $f\in\mci\sigma$ such that $T_a\canbe_\mu fa$ for
  every $a\in\mci\tau$.  Thus $T\canbe_\sigma f$.

  \br $\sigma=\tau\mu,~{\Leftarrow}$.  Let
  $T\canbe_\sigma f$ and $s,t\in T$.  We show
  $s\comp_\sigma t$.  Let $u\comp_\tau v$.  It suffices
  to show $su\comp_\mu tv$.  By the inductive hypothesis
  $u,v\canbe_\tau a$ for some value $a$.  Hence
  $su,tv\canbe_\mu fa$.  Thus $su\comp_\mu tv$ by
  the inductive hypothesis.
\end{proof}

\begin{lem}[Admissibility]
  \label{lemma-inhabitation}
  For every variable $x:\sigma$ there is some $a\in\mcd\sigma$
  such that $x\canbe a$.  In particular,
  $\mcd\sigma$ is a nonempty set for every type
  $\sigma$.
\end{lem}

\begin{proof}
  Let $x:\sigma$ be a variable.  By
  Lemma~\ref{lem-compatibility}\,(2) we know $x\comp_\sigma x$.
  Hence $\set{x}$ is compatible.  By
  Lemma~\ref{lem-common-value} there exists a value $a$
  such that $x\canbe_\sigma a$.  The claim follows
  since $a\in\mcd\sigma$ by definition of $\mcd\sigma$.
\end{proof}

\begin{lem}[Functionality]
  \label{lem-functionality}
  If $s\canbe_\sigma a$, $t\canbe_\sigma b$, and
  $(s{=}t)\in E$ , then $a=b$.
\end{lem}

\begin{proof}
  By contradiction and induction on $\sigma$.  Assume
  $s\canbe_\sigma a$, $t\canbe_\sigma b$, $(s{=}t)\in
  E$, and $a\neq b$.  Case analysis.

  $\sigma=o$.  By $\EBQ$ either $s,t\in E$ or $\neg
  s,\neg t\in E$.  Hence $a$ and $b$ are either both
  $1$ or both $0$.  Contradiction.

  $\sigma=\alpha$.  Since $a\neq b$, there must be
  discriminating terms of type $\alpha$.  Since
  $(s{=}t)\in E$, we know by N3 and \ECon that $s$ and
  $t$ are normal and discriminating.  Hence $s\in a$
  and $t\in b$.  Contradiction by
  Proposition~\ref{prop-diff-discs}\,(2).

  $\sigma = \tau\mu$.  Since $a\neq b$, there is some
  $c\in\mcd\tau$ such that $ac\not= bc$.
  By the
  definition of $\mcd\tau$ and
  Lemma~\ref{prop-norm-poss-value} there is a normal
  term $u$ such that $u\canbe_\tau c$.
  Hence $su\canbe ac$ and $tu\canbe bc$.  By
  Proposition~\ref{prop-norm-poss-value} $\nf{su}\canbe_\mu
  ac$ and $\nf{tu}\canbe_\mu bc$.
  By $\EFQ$ the
  equation $\nf{su} = \nf{tu}$ is in~$E$.
  Contradiction by the inductive hypothesis.
\end{proof}

We now define the canonical interpretations for the logical constants:
\begin{align*}
  \N{\mcl({\neg})}&\eqdef\lam{a{\in}\mcd o}{~\Cond{a{=}1}01}\\ 
  \N{\mcl({=_\sigma})}&\eqdef\lam{a{\in}\mcd\sigma}{~\lam{b{\in}\mcd\sigma}{~\Cond{a{=}b}10}}
\end{align*}

\begin{lem}[Logical Constants]
  \label{lem-log-constants}
  $c\canbe\mcl(c)$ for every logical constant $c$.
\end{lem}

\begin{proof}
  We show $\neg\canbe \mcl(\neg)$ by contradiction.
  Let $s\canbe_oa$ and assume $\neg s \ncanbe\mcl(\neg) a$. Case analysis.
  \begin{enumerate}[$\bullet$]
  \item $a=0$.  Then $\nf{s}\notin
    E$ and $\neg\nf{\neg s}\in E$.  Contradiction
    by N3 and~$\EDN$.
  \item $a=1$.  Then $\neg\nf{s}\notin E$ and $\nf{\neg s}\in E$.
    Contradiction by N3.
  \end{enumerate}
  Finally, we show $(=_\sigma)\canbe\mcl(=_\sigma)$ by
  contradiction.  Let $s\canbe_\sigma a$,
  $t\canbe_\sigma b$, and
  $(s{=_\sigma}t)\ncanbe\mcl(=_\sigma)ab$.  Case analysis.
  \begin{enumerate}[$\bullet$]
  \item $a=b$.  Then $\nf{s}\notq\nf{t}$ by N3 and
    $s,t\canbe a$.  Thus $s\comp t$ by
    Lemma~\ref{lem-common-value}.  Contradiction by
    Lemma~\ref{lem-compatibility}\,(1).
  \item $a\neq b$.  Then $(\nf{s}{=}\nf{t})\in E$ by
    N3. Hence $a=b$ by
    Proposition~\ref{prop-norm-poss-value}
    and Lemma~\ref{lem-functionality}.  Contradiction.\qedhere
  \end{enumerate}
\end{proof}

\begin{thm}[Model Existence]
  \label{theo-model-exist}
  Every evident branch is satisfiable.  Moreover, every
  complete evident branch has a surjective model, and
  every finite evident branch has a finite model.
\end{thm}

\begin{proof}
  Let $E$ be an evident branch and $\canbe$ be the
  value system for $E$.  By
  Proposition~\ref{prop-mod-ex-o},
  Lemma~\ref{lemma-inhabitation}, and
  Lemma~\ref{lem-log-constants} we have a logical
  interpretation $\mci$ that is admissible for
  $\canbe$.  By Lemma~\ref{lemma-adm-model} $\mci$ is a
  model of $E$.  

  Let $E$ be complete.  By
  Theorem~\ref{theo-admissible-interpretations} we know
  that $\mci$ is surjective if $\canbe$ is functional.
  Let $s\canbe_\beta a$ and $s\canbe_\beta b$.  We show
  $a=b$.  By Proposition~\ref{prop-norm-poss-value} we
  can assume that $s$ is normal.  Thus $s{=}s$ is
  normal by N3. Since $\mci$ is a model of $E$, we know
  that the formula $s{\neq}s$ is not in $E$.  Since $E$
  is complete, we know that ${s}{=}{s}$ is
  in~$E$.  By Lemma~\ref{lem-functionality} we have
  $a=b$.

  If $E$ is finite, $\mci\alpha=\mcd\alpha$ is finite
  by Proposition~\ref{prop-finite-discs}.
\end{proof}

\section{Abstract Consistency}

We now extend the model existence result for evident
branches to abstract consistency classes, following the
corresponding development for first-order
logic~\cite{SmullyanBook}.
Notions of abstract consistency for simple type theory have
been previously considered in~\cite{Andrews71,Kohlhase93a,KohlhaseTableaux1995,Benzmuller99a,BenzKoh98,BBK04,BBKweb04,Brown2004a,BrownARHO}.
Equality was treated as Leibniz equality in~\cite{Andrews71}.
Abstract consistency conditions for
primitive equality corresponding to reflexivity and substutivity properties
were given by Benzm\"uller in~\cite{Benzmuller99a,Benzmuller99b}. 
A primitive identity predicate $=_\sigma$ was considered in~\cite{BBK04}
but the abstract consistency conditions for $=_\sigma$ essentially reduced it
to Leibniz equality.  
Conditions for $=_\sigma$ analogous to $\ACon$ first appeared in~\cite{Brown2004a}.

An \emph{abstract consistency class} is a set $\Gamma$
of branches such that every branch $A\in\Gamma$
satisfies the conditions in
Figure~\ref{fig:abs-consistency}.  An abstract
consistency class $\Gamma$ is \emph{complete} if for
every branch $A\in\Gamma$ and every normal formula $s$
either $A\cup\set{{s}}$ or $A\cup\set{\neg{s}}$ is
in~$\Gamma$.  
The completeness condition was called ``saturation'' in~\cite{BBK04}.
As discussed in~\cite{BBK2009} and the conclusion of~\cite{BBK04},
the condition corresponds to having a cut rule in a calculus.
In~\cite{BBKweb04} conditions analogous to $\ADec$ and $\AMat$ appear (using Leibniz equality)
and a model existence theorem is proven with these conditions replacing saturation.
The use of Leibniz equality means that there was still not a cut-free treatment of equality in~\cite{BBKweb04}.

\begin{figure}[tp]
  \renewcommand{\arraystretch}{1.4}
  \begin{tabular}{c>{\raggedright}p{120mm}}
    \emph{\ADN}&If $\neg\neg s$ is in $A$, 
    then $A\cup\set{s}$ is in $\Gamma$.
   \tabularnewline
    \emph{\ABQ}&If $s =_o t$ is in $A$, 
    then either $A\cup\set{s,t}$
    or $A\cup\set{\neg s,\neg t}$ is in $\Gamma$.
    \tabularnewline
    \emph{\ABE}&If $s\neq_o t$ is in $A$, 
    then either $A\cup\set{s,\neg t}$
    or $A\cup\set{\neg s,t}$ is in $\Gamma$.
    \tabularnewline
    \emph{\AFQ}&If $s =_{\sigma\tau} t$ is in $A$,\\ 
    then $A\cup\set{\nf{su}\neq\nf{tu}}$ is in $\Gamma$ for every normal~$u:\sigma$.
    \tabularnewline
    \emph{\AFE}&If $s\neq_{\sigma\tau} t$ is in $A$, 
    then $A\cup\set{\nf{sx}\neq\nf{tx}}$ is in $\Gamma$ for some variable $x$.
    \tabularnewline
    \emph{\AMat}&If $x\ddd s n$ is in $A$ and $\neg x\ddd t n$ is in $A$,\\
    then $n\geq 1$ and $A\cup\set{s_i\neq t_i}$ is in $\Gamma$ for some $i\in\set{1\cld n}$.
    \tabularnewline
    \emph{\ADec}&If $x\ddd s n\neq_\alpha x\ddd t n$ is in $A$,
    then $n\ge 1$ and $A\cup\set{s_i\neq t_i}$ is in $\Gamma$ for some $i\in\set{1\cld n}$.
    \tabularnewline
    \emph{\ACon}&If $s=_\alpha t$ and $u \neq_\alpha v$  are in $A$,\\
    then either $A\cup\set{s\neq u,t\neq u}$
    or $A\cup\set{s\neq v,t\neq v}$ is in $\Gamma$.
  \end{tabular}
  \caption{Abstract consistency conditions (must hold for every $A\in\Gamma$)}
  \label{fig:abs-consistency}
\end{figure}

\begin{prop}
  Let $A$ be a branch.  Then $A$ is evident if and only
  if $\set{A}$ is an abstract consistency class.
  Moreover, $A$ is a complete evident branch if and
  only if $\set{A}$ is a complete abstract consistency
  class.
\end{prop}

\begin{lem}[Extension Lemma]
  \label{lem:extension}
  Let $\Gamma$ be an abstract consistency class and
  $A\in\Gamma$.  Then there exists an evident branch
  $E$ such that $A\subseteq E$.  Moreover, if $\Gamma$
  is complete, a complete evident branch $E$ exists
  such that $A\subseteq E$.
\end{lem}

\begin{proof}
  Let $u_0,u_1,u_2,\ldots$ be an enumeration of all
  normal formulas.  We construct a sequence
  $A_0\subseteq A_1 \subseteq A_2 \subseteq \cdots$ of
  branches such that every $A_n\in\Gamma$.  Let $A_0
  \deq A$.  We define $A_{n+1}$ by cases.  If there is
  no $B\in\Gamma$ such that $A_{n}\cup\{u_n\}\subseteq
  B$, then let $A_{n+1}\deq A_n$.  Otherwise, choose
  some $B\in\Gamma$ such that
  $A_{n}\cup\{u_n\}\subseteq B$.  We consider two
  subcases.
  \begin{enumerate}[(1)]
  \item If $u_n$ is of the form $s\neq_{\sigma\tau} t$,
    then choose $A_{n+1}$ to be
    $B\cup\{\nf{sx}\neq\nf{tx}\}\in\Gamma$ for some
    variable $x$.  This is possible since $\Gamma$
    satisfies $\AFE$.
  \item If $u_n$ is not of this form, then let
    $A_{n+1}$ be $B$.
  \end{enumerate}
  Let $\displaystyle E:=\bigcup_{n\in\NN} A_n$. 
We show that $E$ satisfies the evidence conditions.
\begin{enumerate}[\EMat]
\item[{\EDN}] Assume $\neg\neg s$ is in $E$.
  Let $n$ be such that $u_n=s$.  Let
  $r\geq n$ be such that $\neg \neg s$ is in $A_r$.
  By $\ADN$,
  $A_r\cup\{s\}\in\Gamma$.  Since
  $A_n\cup\{s\}\subseteq A_r\cup\{s\}$, we have $s\in
  A_{n+1}\subseteq E$.
\item[{\EMat}] Assume $x\ddd s n$ and $\neg x\ddd
  t n$ are in $E$.  For each
  $i\in\set{1\cld n}$, let $m_i$ be such that $u_{m_i}$
  is $s_i\neq t_i$.  Let $r \ge m_1,\ldots,m_n$ be such
  that $x\ddd s n$ and $\neg x\ddd t n$ are in $A_r$.
  By $\AMat$ $n\geq 1$ and there is some $i\in\set{1\cld n}$ such
  that $A_r\cup\{s_i\neq t_i\}\in\Gamma$.  Since
  $A_{m_i}\cup\{s_i\neq t_i\}\subseteq A_r\cup\{s_i\neq
  t_i\}$, we have $(s_i\neq t_i)\in A_{m_{i}+1}\subseteq
  E$.
\item[{\EDec}] Similar to $\EMat$  
\item[{\ECon}] Assume $s=_\alpha t$ and $u
  \neq_\alpha v$ are in $E$.  Let $n,m,j,k$ be such that
  $u_n$ is $s\neq u$, $u_m$ is $t\neq u$, $u_j$ is
  $s\neq v$ and $u_k$ is $t\neq v$.  Let $r\ge n,m,j,k$
  be such that $s=_\alpha t$ and $u \neq_\alpha v$ are in
  $A_r$.  By $\ACon$ either $A_r\cup\{s\neq u,t\neq
  u\}$ or $A_r\cup\{s\neq v,t\neq v\}$ is in $\Gamma$.
  Assume $A_r\cup\{s\neq u,t\neq u\}$ is in $\Gamma$.
  Since $A_n\cup\{s\neq u\}\subseteq A_r\cup\{s\neq
  u,t\neq u\}$, we have $s\neq u\in A_{n+1}\subseteq
  E$.  Since $A_m\cup\{t\neq u\}\subseteq
  A_r\cup\{s\neq u,t\neq u\}$, we have $t\neq u\in
  A_{m+1}\subseteq E$.  Next assume $A_r\cup\{s\neq
  v,t\neq v\}$ is in $\Gamma$.
  By a similar argument we know $s\neq v$ and $t\neq v$ must be in $E$.
\item[{\EBQ}] Assume $s =_o t$ is in $E$.  Let
  $n,m,j,k$ be such that $u_n=s$, $u_m=t$, $u_j=\neg s$
  and $u_k=\neg t$.  Let $r\ge n,m,j,k$ be such that
  $s =_o t$ is in $A_r$.  By $\ABQ$ either
  $A_r\cup\{s,t\}$ or $A_r\cup\{\neg s,\neg t\}$ is in
  $\Gamma$.  Assume $A_r\cup\{s,t\}$ is in
  $\Gamma$.  Since $A_n\cup\{s\}\subseteq
  A_r\cup\{s,t\}$, we have $s\in E$.  Since
  $A_m\cup\{t\}\subseteq A_r\cup\{s,t\}$, we
  have $t\in E$.  Next assume $A_r\cup\{\neg
  s,\neg t\}$ is in $\Gamma$.  Since $A_j\cup\{\neg
  s\}\subseteq A_r\cup\{\neg s, \neg t\}$, we have $\neg
  s\in E$.  Since $A_k\cup\{\neg t\}\subseteq A_r\cup\{\neg
  s,\neg t\}$, we have $\neg t\in E$.
\item[{\EBE}] Similar to $\EBQ$ 
\item[{\EFQ}] Assume $s =_{\sigma\tau} t$ is in $E$ and
  $u:\sigma$ is normal.  Let $n$ be such that
  $u_n$ is $\nf{su} =_\tau \nf{tu}$.
  Let $r\ge n$ be
  such that $s =_{\sigma\tau} t$ is in $A_r$.
  By $\AFQ$ we know $A_r\cup\{\nf{su} =_\tau \nf{tu}\}$ is
  in $\Gamma$.  Hence $\nf{su} =_\tau\nf{tu}$ is in
  $A_{n+1}$ and also in $E$.
\item[{\EFE}] Assume $s\neq_{\sigma\tau} t$ is in
  $E$.  Let $n$ be such that $u_n$ is
  $s\neq_{\sigma\tau} t$.  Let $r\geq n$ be such that
  $s\neq_{\sigma\tau} t$ is in $A_r$.  Since
  $A_n\cup\{u_n\}\subseteq A_r$, there is some
  variable $x$ such that $\nf{sx}\neq_\tau\nf{tx}$ is
  in $A_{n+1}\subseteq E$.
\end{enumerate}
It remains to show that~$E$ is complete if $\Gamma$ is
complete.  Let $\Gamma$ be complete and $s$ be a normal
formula.  We show that ${s}$ or $\neg{s}$ is in $E$.
Let $m$, $n$ be such that $u_m={s}$ and $u_n=\neg{s}$.
We consider $m<n$. (The case $m>n$ is symmetric.)  If
${s}\in A_{n}$, we have ${s}\in E$.  If ${s}\notin
A_{n}$, then $A_n\cup\set{{s}}$ is not in $\Gamma$.
Hence $A_n\cup\set{\neg{s}}$ is in $\Gamma$ since
$\Gamma$ is complete.  Hence $\neg{s}\in A_{n+1}\incl
E$.
\end{proof}

\begin{thm}[Model Existence]
  \label{theo-acc-model-existence}
  Every member of an abstract consistency class has a
  model, which is surjective if the consistency class
  is complete.
\end{thm}

\begin{proof}
  Let $A\in\Gamma$ where $\Gamma$ is an abstract
  consistency class.  By Lemma~\ref{lem:extension} we
  have an evident set $E$ such that $A\incl E$, where
  $E$ is complete if $\Gamma$ is complete.  The claim
  follows with Theorem~\ref{theo-model-exist}.
\end{proof}

\section{Completeness}
\label{sec:completeness}

It is now straightforward to prove the completeness of
the tableau calculus $\TS$. 
Let~\emph{$\GammaT$} be the
set of all finite branches that are not refutable.


\begin{lem}
  \label{lem:acc-completeness}
  $\GammaT$ is an abstract consistency class.
\end{lem}

\proof
    We have to show that $\GammaT$ satisfies the
    abstract consistency conditions.
\begin{enumerate}[\AMat]
\item[{\ADN}] Assume $\neg\neg s$ is in $A$ and
  $A\cup\{s\}\notin\GammaT$.  Then we can refute $A$
  using $\TRDN$.
\item[{\AMat}] Assume $\{x\ddd s n,\neg x\ddd t
  n\}\subseteq A$ and $A\cup\{s_i\neq
  t_i\}\notin\GammaT$ for all $i\in\set{1\cld n}$.
  Then we can refute $A$ using \TRMat.
\item[{\ADec}] Assume $x\ddd s n\neq_\alpha x\ddd t
  n$ is in $A$ and $A\cup\{s_i\neq t_i\}\notin\GammaT$
  for all $i\in\set{1\cld n}$.
  Then we can refute $A$ using \TRDec.
\item[{\ACon}] Assume $s=_\alpha t $ and
  $u\neq_\alpha v$ are in $A$ but $A\cup\{s\neq u,t\neq
  u\}$ and $A\cup{\set{s\neq v,t\neq v}}$ are not in
  $\GammaT$.  Then we can refute $A$ using \TRCon.
\item[{\ABQ}] Assume $s=_o t$ is in $A$,
  $A\cup\{s,t\}\notin\GammaT$ and $A\cup\{\neg s,\neg t\}\notin\GammaT$.
  Then we can refute $A$ using
  \TRBQ.
\item[{\ABE}] Assume $s\neq_o t$ is in $A$,
  $A\cup\{s,\neg t\}\notin\GammaT$ and $A\cup\{\neg
  s,t\}\notin\GammaT$.  Then we can refute $A$ using
  \TRBE.
\item[{\AFQ}] Let $(s=_{\sigma\tau}t)\in A\in\GammaT$.
  Suppose $A\cup\set{\nf{su}{=}\nf{tu}}\notin\GammaT$
  for some normal $u\in\Wff_\sigma$.  Then
  $A\cup\set{\nf{su}{=}\nf{tu}}$ is refutable and so
  $A$ is refutable by $\TRFQ$.
\item[{\AFE}] Let $(s{\neq}_{\sigma\tau}t)\in
  A\in\GammaT$.  Suppose
  $A\cup\set{\nf{sx}{\neq}\nf{tx}}\notin\GammaT$ for
  every variable $x:\sigma$.  Then
  $A\cup\set{\nf{sx}{\neq}\nf{tx}}$ is refutable for
  every $x:\sigma$.  Hence $A$ is refutable using~\TRFE
  and the finiteness of $A$.  Contradiction.\qed
\end{enumerate}

\begin{thm}[Completeness]
  \label{thm:completeness}
  Every unsatisfiable finite branch is refutable.
\end{thm}

\begin{proof}
  By contradiction.  Let $A$ be an unsatisfiable finite
  branch that is not refutable.  Then $A\in\GammaT$ and
  hence $A$ is satisfiable by
  Lemma~\ref{lem:acc-completeness} and
  Theorem~\ref{theo-acc-model-existence}.  Contradiction.
\end{proof}

\section{Compactness and Countable Models}

It is known~\cite{Henkin50,AndrewsBook} that
simple type theory is compact and has the
countable-model property.  We use the opportunity and
show how these properties follow with the results we
already have.  It is only for the existence of countable
models that we make use of complete evident sets and
complete abstract consistency classes.

A branch $A$ is \emph{sufficiently pure} if for every
type $\sigma$ there are infinitely many variables of
type $\sigma$ that do not occur free in the formulas of $A$.
Let $\Gammacomp$ be the set of all sufficiently pure
branches $A$ such that every finite subset of $A$ is
satisfiable.  We write \emph{$\fsubseteq$} for the
finite subset relation.

\begin{lem}
  \label{lem-aux-compactness}
  Let $A\in\Gammacomp$ and $\dd{B}n$ be finite branches
  such that $A\cup B_i\notin\Gammacomp$ for all
  $i\in\set{1\cld n}$.  Then there exists a finite
  branch $A'\fsubseteq A$ such that $A'\cup B_i$ is
  unsatisfiable for all $i\in\set{1\cld n}$.
\end{lem}

\begin{proof}
  By the assumption, we have for every $i\in\set{1\cld
    n}$ a finite and unsatisfiable branch $C_i\incl
  A\cup B_i$.  The branch $A':=(C_1\cup\dots\cup
  C_n)\cap A$ satisfies the claim.
\end{proof}

\begin{lem}
  \label{lem:acc-compactness}
  $\Gammacomp$ is a complete abstract consistency
  class.
\end{lem}

\begin{proof}
  We verify the abstract consistency conditions using
  Lemma~\ref{lem-aux-compactness} tacitly.
\begin{enumerate}[\AMat]
\item[{\ADN}] Assume $\neg\neg s$ is in $A$ and
  $A\cup\{s\}\notin\Gammacomp$.  There is some
  $A'\fsubseteq A$ such that $A'\cup\{s\}$ is
  unsatisfiable.  There is a model of
  $A'\cup\{\neg \neg s\}\fsubseteq A$.
  This is also a model of
  $A'\cup\{s\}$, contradicting our choice of $A'$.
\item[{\AMat}] Assume $x\ddd s n$ and $\neg x\ddd
  t n$ are in $A$ and $A\cup\{s_i\neq
  t_i\}\notin\Gammacomp$ for all $i\in\set{1\cld n}$.
  There is some $A'\fsubseteq A$ such that
  $A'\cup\{s_i\neq t_i\}$ is unsatisfiable for all
  $i\in\set{1\cld n}$.  There is a model $\mci$ of
  $A'\cup\{x\ddd s n,\neg x\ddd t n\}\fsubseteq A$.
  Since $\hat\mci (x\ddd s n) \neq \hat\mci (x\ddd t
  n)$, we must have $\hat\mci(s_i) \neq \hat\mci(t_i)$
  for some $i\in\set{1\cld n}$ (and in particular $n$ must not be $0$).  Thus $\mci$ models
  $A'\cup\{s_i\neq t_i\}$, contradicting our choice
  of~$A'$.
\item[{\ADec}] Similar to $\AMat$ 
\item[{\ACon}] Assume $s=_\alpha t $ and $u\neq_\alpha v$
  are in $A$, $A\cup\{s\neq u,t\neq
  u\}\notin\Gammacomp$ and $A\cup\{{s\neq v},t\neq
  v\}\notin\Gammacomp$.  There is some $A'\fsubseteq A$
  such that $A'\cup\{{s\neq u},t\neq u\}$ and
  $A'\cup\{s\neq v,t\neq v\}$ are unsatisfiable.  There
  is a model $\mci$ of $A'\cup\{{s = t},{u\neq
  v}\}\fsubseteq A$.  Since $\hat\mci(s) = \hat\mci(t)$
  and $\hat\mci(u) \neq \hat\mci(v)$, we either have
  $\hat\mci(s) \neq \hat\mci(u)$ and $\hat\mci(t) \neq
  \hat\mci(u)$ or $\hat\mci(s) \neq \hat\mci(v)$ and
  $\hat\mci(t) \neq \hat\mci(v)$.  Hence $\mci$ models
  either $A'\cup\{s\neq u,t\neq u\}$ or $A'\cup\{s\neq
  v,t\neq v\}$, contradicting our choice of $A'$.
\item[{\ABQ}] Assume $s =_o t$ is in $A$,
  $A\cup\{s, t\}\notin\Gammacomp$ and $A\cup\{\neg
  s,\neg t\}\notin\Gammacomp$.  There is some $A'\fsubseteq
  A$ such that $A'\cup\{s,t\}$ and $A'\cup\{\neg
  s,\neg t\}$ are unsatisfiable.  There is a model of
  $A'\cup\{s =_o t\}\fsubseteq A$.  This is also a
  model of $A'\cup\{s, t\}$ or $A'\cup\{\neg
  s,\neg t\}$.
\item[{\ABE}] Assume $s\neq_o t$ is in $A$,
  $A\cup\{s,\neg t\}\notin\Gammacomp$ and $A\cup\{\neg
  s,t\}\notin\Gammacomp$.  There is some $A'\fsubseteq
  A$ such that $A'\cup\{s,\neg t\}$ and $A'\cup\{\neg
  s,t\}$ are unsatisfiable.  There is a model of
  $A'\cup\{s\neq_o t\}\fsubseteq A$.  This is also a
  model of $A'\cup\{s,\neg t\}$ or $A'\cup\{\neg
  s,t\}$.
\item[{\AFQ}] Assume $s =_{\sigma\tau} t $ is in $A$
  but $A\cup \{\nf{su} =_\tau\nf{tu}\}$ is not in
  $\Gammacomp$ for some normal $u\in\Wff_\sigma$.
  There is some $A'\fsubseteq A$ such that
  $A'\cup\{\nf{su}=\nf{tu}\}$ is unsatisfiable.  There
  is a model $\mci$ of $A'\cup\{s = t\}\fsubseteq A$.
  Since $\hat\mci(s) =\hat\mci(t)$, we know
  $\hat\mci(\nf{su}) = \hat\mci(su) =
  \hat\mci(s)\hat\mci(u) = \hat\mci(t)\hat\mci(u) =
  \hat\mci(tu) = \hat\mci(\nf{tu})$ using N4.  Hence
  $\mci$ is a model of $A'\cup\{\nf{su}=\nf{tu}\}$, a
  contradiction.  
\item[{\AFE}] Assume $s\neq_{\sigma\tau} t $ is in $A$.
  Since $A$ is sufficiently pure, there is a variable
  $x:\sigma$ which does not occur in $A$.  Assume
  $A\cup\{\nf{sx}\neq\nf{tx}\}\notin\Gammacomp$.  There
  is some $A'\fsubseteq A$ such that
  $A'\cup\{\nf{sx}\neq\nf{tx}\}$ is unsatisfiable.
  There is a model $\mci$ of $A'\cup\{s\neq
  t\}\fsubseteq A$.  Since
  $\hat\mci(s)\neq\hat\mci(t)$, there must be some
  $a\in\mci\sigma$ such that
  $\hat\mci(s)a\neq\hat\mci(t)a$.  Since $x$ does not
  occur free in $A$, we know
  $\widehat{\mci^x_a}(sx)\neq\widehat{\mci^x_a}(tx)$ and
  $\mci^x_a$ is a model of $A'$.  Since
  $\widehat{\mci^x_a}(\nf{sx}) =\widehat{\mci^x_a}(sx)$
  and $\widehat{\mci^x_a}(\nf{tx}) =\widehat{\mci^x_a}(tx)$ by N4, we
  conclude $\mci^x_a$ is a model of
  $A'\cup\{\nf{sx}\neq\nf{tx}\}$, contradicting our
  choice of $A'$.
\end{enumerate}
We show the completeness of $\Gammacomp$
by contradiction.  Let $A\in\Gammacomp$ and $s$ be a
normal formula such that $A\cup\set{{s}}$ and
$A\cup\set{\neg{s}}$ are not in $\Gammacomp$.  Then
there exists $A'\fsubseteq A$ such that
$A'\cup\set{{s}}$ and $A'\cup\set{\neg{s}}$ are
unsatisfiable.  Contradiction since~$A'$ is
satisfiable.
\end{proof}

\begin{thm}\label{thm:compactness-ls}
  Let $A$ be a branch such that every finite subset of
  $A$ is satisfiable.  Then $A$ has a countable model.
\end{thm}

\begin{proof}
  Without loss of generality we assume $A$ is
  sufficiently pure.  Then $A\in\Gammacomp$.  Hence $A$
  has a countable model by
  Lemma~\ref{lem:acc-compactness} and
  Theorem~\ref{theo-acc-model-existence}.
\end{proof}

\section{EFO Fragment}


We now turn to the EFO fragment of STT as first reported in~\cite{BrownSmolkaEFO}.
The EFO fragment contains first-order logic
and enjoys the usual properties of first-order logic.
We will show completeness and compactness with respect to standard models.
We will also prove that countable models for evident EFO sets exist.

Suppose STT were given with $\neg$, $\limplies$, $=_\sigma$ and $\forall_{\!\sigma}$.
Then the natural definition of EFO would restrict
$=_\sigma$ and $\forall_{\!\sigma}$ to the case where $\sigma$ is a base type.
To avoid redundancy our definition of EFO will also exclude the case where $\sigma = o$.

Our definition of EFO assumes the logical constants $\neg:oo$, $\limplies:ooo$,
$=_\alpha:\alpha\alpha o$ and $\forall_{\!\alpha}:(\alpha o)o$ where
$\alpha$ ranges over sorts.  We call these constants \emph{EFO constants}.
For an assignment to be logical we require that it interprets the logical
constants as usual.  In particular, $\mci(\forall_{\!\alpha})$ must be
the function returning $1$ iff its argument is the constant $1$ function.

We say a term is \emph{EFO} if it only contains the logical constants $\neg$, $\to$, $=_\alpha$ and $\forall_{\!\alpha}$.
Let \emph{$\EFO_\sigma$} be the set of EFO terms of type $\sigma$.
A term is \emph{quasi-EFO} if it is EFO or of the form $s\not=_{\sigma} t$ where $s,t$ are EFO
and $\sigma$ is a type. 
A branch $E$ is \emph{EFO} if every member of $E$ is quasi-EFO.
The example tableau shown in Figure~\ref{fig:refutation} only contains EFO branches.

\begin{figure}
\begin{mathpar}
  \inferrule*[left=\emph{\TRFDN}~]{\neg\neg s}{s}
  \and
  \inferrule*[left=\emph{\TRFBE}~]{s\neq_ot}{s\,,\,\neg t~\mid~\neg s\,,\,t}
  \and
  \inferrule*[left=\emph{\TRFImp}~]{s\limplies t}{\neg s\mid t}
  \and
  \inferrule*[left=\emph{\TRFImpN}~]{\neg(s\limplies t)}{s\,,\,\neg t}
  \\
  \inferrule*[left=\emph{\TRFMat}~,right=~$n\geq 0$] {xs_1\dots
    s_n\,,\,\neg xt_1\dots t_n} {s_1\neq t_1\mid\dots\mid s_n\neq t_n}
  \and
  \inferrule*[left=\emph{\TRFDec}~,right=~$n\geq 0$] {xs_1\dots
    s_n\neq_\alpha xt_1\dots t_n} {s_1\neq t_1\mid\dots\mid s_n\neq t_n}
  \\
  \inferrule*[left=\emph{\TRFFE}~,right=~$x:\sigma$ fresh]
  {s\neq_{\sigma\tau} t}{\nf{sx}\neq\nf{tx}}
  \and
  \inferrule*[left=\emph{\TRFCon}~]
  {s=_\alpha t\,,\,u\neq_\alpha v}
  {s\neq u\,,\,t\neq u\mid s\neq v\,,\,t\neq v}
  \\
  \inferrule*[left=\emph{\TRFall}~,right=~$u\in\EFO_\alpha$ normal]
  {\forall_{\!\alpha} s}{\nf{su}}
  \and
  \inferrule*[left=\emph{\TRFalln}~,right=~$x:\alpha$ fresh]
  {\neg\forall_{\!\alpha} s}{\neg\nf{sx}}
\end{mathpar}
\caption{Tableau rules for EFO}
\label{fig:rulesefo}
\end{figure}
The tableau rules in Figure~\ref{fig:rulesefo} define a tableau calculus $\TSF$
for EFO branches
up to restrictions on applicability given in Section~\ref{sec:efo-complete}.
After showing a model existence theorem,
we will precisely define the tableau calculus $\TSF$ 
and prove it is complete for EFO branches.
The completeness result will be with respect to standard models.
For some fragments of EFO
the tableau calculus $\TSF$ will terminate, yielding decidability results.

\section{EFO Evidence and Compatibility}

We say an EFO branch $E$ is evident if it satisfies the evidence conditions in Figure~\ref{fig:evidence}
and the following additional conditions.\\
  \renewcommand{\arraystretch}{1.4}
  \begin{tabular}{c>{\raggedright}p{120mm}}
    \emph{\EImp}&If $s\limplies t$ is in $E$, then $\neg s$ or $t$ is in $E$.
    \tabularnewline
    \emph{\EImpN}&If $\neg(s\limplies t)$ is in $E$, then $s$ and $\neg t$ are in $E$.
    \tabularnewline
    \emph{\Eall}&If $\forall_{\!\alpha} s$ is in $E$, 
    then $\nf{su}$ is in $E$ for every $\alpha$-discriminating $u$ in $E$.
    \tabularnewline
    \emph{\Ealld}&If $\forall_{\!\alpha} s$ is in $E$,
    then $\nf{su}$ is in $E$ for some normal EFO term $u:\alpha$.  
    \tabularnewline
    \emph{\Ealln}&If $\neg \forall_{\!\alpha} s$ is in $E$, 
    then $\neg\nf{sx}$ is in $E$ for some variable $x$.
  \end{tabular}\\[2mm]
We say an EFO branch $E$ is \emph{EFO-complete} if for all normal $s\in\EFO_o$ either $s\in E$ or $\neg s\in E$.

The condition $\Eall$ is the usual condition for universal quantifiers with instantiations restricted
to $\alpha$-discriminating terms.  Since there may be no $\alpha$-discriminating terms in $E$,
we also include the condition $\Ealld$ to ensure that at least one instantiation has been made.
Without the condition $\Ealld$, the set $\{\forall_{\!\alpha} x.\neg(y\limplies y)\}$ would be evident.

Let $E$ be an evident EFO branch. 
Compatibility can be defined exactly as in Section~\ref{sec:compat}
and Lemma~\ref{lem-compatibility} holds.
In the proof of Lemma~\ref{lem-efo-log-constants} below, we will need
to know that if $E$ has some $\alpha$-discriminating term,
then all $\alpha$-discriminants are nonempty.  Since $\alpha$-discriminants are
maximal sets of $\alpha$-discriminating terms, it is enough to prove
every $\alpha$-discriminating term is compatible with itself.
To be concrete, we must prove $s\not=_\alpha s$ is never in $E$.
One way we could ensure this is to include it as an evidence condition
and have a corresponding tableau rule of the form:
\begin{mathpar}
  \inferrule*[left=\emph{\TRFBotD}~]{s\neq_\alpha s}{\,}
\end{mathpar}
This was the choice taken in~\cite{BrownSmolkaEFO}.
One drawback to including the rule $\TRFBotD$ in the ground calculus is that a lifting lemma
will be more difficult to show when one passes to a calculus with variables.

Another alternative is to remove the restriction on instantiations in the
rule $\TRFall$.  If we do not restrict $\TRFall$ to discriminating terms,
then we can show the existence of a model without knowing a priori
that $\alpha$-discriminants are nonempty in the presence of $\alpha$-discriminating terms.

In order to obtain a strong completeness result,
we will not follow either of these alternatives.
Instead we prove that all terms are compatible with themselves.
First we prove EFO constants are compatible with themselves.

\begin{lem}
  \label{efoconst-compat}
  For every EFO constant $c$, $c\comp c$.
\end{lem}
\begin{proof}
  Case analysis.
  $\neg\comp\neg$ follows from N3 and $\EDN$.
  $\to\comp\to$ follows from N3, $\EImp$ and $\EImpN$.
  $=_\alpha\comp =_\alpha$ follows from N3 and $\ECon$.
  We show $\forall_{\!\alpha}\comp \forall_{\!\alpha}$.
  Let $s \comp_{\alpha o} t$ be given.  Assume $\forall s \ncomp \forall t$.
  Without loss of generality, assume $\nf{\forall s}$ and $\neg\nf{\forall t}$ are in $E$.
  By $\Ealln$ we have $\neg \nf{tx}$ in $E$ for some variable $x:\alpha$.
  By $\Ealld$ we have $\nf{su}$ in $E$ for some normal EFO term $u$.
  Since $su \ncomp_o tx$, we must have $u\ncomp_\alpha x$.  In particular,
  $x$ must be an $\alpha$-discriminating term.
  By $\Eall$ we have $\nf{sx}$ is in $E$.
  Hence we must have $x\ncomp_\alpha x$, contradicting Lemma \ref{lem-compatibility}\,(2).
\end{proof}

Next we prove compatibility respects normalization.
\begin{lem}\label{lem-norm-comp}
  For all $s,t:\sigma$, $s\comp_\sigma t$ iff $\nf{s}\comp_\sigma\nf{t}$.
\end{lem}
\begin{proof}
  Induction on types.  At base types this follows from N1 and the definition of compatibility.
  Assume $\sigma$ is $\tau\mu$.  Let $u\comp_\tau v$.
  By N2 and the inductive hypothesis (twice) we have
  $su\comp tv$ iff $\nf{su}\comp \nf{tv}$
  iff $\nf{\nf{s}u}\comp \nf{\nf{t}v}$
  iff
  $\nf{s}u\comp \nf{t}v$.
  Hence $s\comp t$ iff $\nf{s}\comp\nf{t}$.
\end{proof}

For two substitutions $\theta$ and $\phi$
we write \emph{$\theta \comp \phi$} when
$\Dom\theta = \Dom\phi$,
$\theta x \comp \phi x$ for every variable $x\in\Dom\theta$
and
$\theta c \comp \phi c$ for every EFO constant $c\in\Dom\theta$.

\begin{lem}
  \label{efotrm-compat-lema}
  For all $s\in\EFO_\sigma$, if $\theta\comp \phi$,
  then $\hat\theta s \comp \hat\phi s$.
\end{lem}
\begin{proof}
  By induction on $s$.  Case analysis.

  \br $s$ is a variable or an EFO constant in $\Dom\theta$.
  The claim follows from $\theta\comp \phi$ and S1.

  \br $s$ is a variable not in $\Dom\theta$.
  The claim follows from S1 and Lemma \ref{lem-compatibility}\,(2).

  \br $s$ is an EFO constant not in $\Dom\theta$.
  The claim follows from S1 and Lemma~\ref{efoconst-compat}.
  
  \br $s=tu$.  By inductive hypothesis $\hat\theta t \comp \hat\phi t$ and $\hat\theta u \comp \hat\phi u$.
  Hence $\hat\theta (tu) \comp \hat\phi (tu)$ using S2.

  \br $s=\lam{x}t$ where $x:\sigma$.  Let $u\comp v$ be given.  We will prove $(\hat\theta s)u\comp (\hat\phi s)v$.
  Using Lemma~\ref{lem-norm-comp} and S3 it is enough to prove
  ${\widehat{\subst\theta{x}{u}}t}\comp{\widehat{\subst\phi{x}{v}}t}$.
  This is the inductive hypothesis with $\subst{\theta}{x}{u}$ and $\subst{\phi}{x}{v}$.
\end{proof}

\begin{lem}
  \label{efotrm-compat-lemb}
  For all $s\in\EFO_\sigma$, $s\comp s$.
\end{lem}
\begin{proof}
  By Lemma~\ref{efotrm-compat-lema} we have
  $\hat{\emptyset}{s}\comp\hat{\emptyset}{s}$.
  We conclude $s\comp s$
  using Lemma~\ref{lem-norm-comp} and S4.
\end{proof}

We can now prove $\alpha$-discriminants are nonempty if $E$ has some $\alpha$-discriminating term.

\begin{lem}
  \label{discr-nonempty}
  If $a$ is an $\alpha$-discriminant and $E$ has an $\alpha$-discriminating term, then
  $a$ is nonempty.
\end{lem}
\begin{proof}
  Let $s$ be $\alpha$-discriminating. We know $s\comp s$ by Lemma~\ref{efotrm-compat-lemb} and so 
  $\{s\}$ is compatible.  If $a$ is empty, then $a\cup\{s\}$ is compatible, contradicting
  maximality of $a$.
\end{proof}

\section{EFO Model Construction}

Let $E$ be an evident EFO branch.  
We inductively define a standard frame $\mcd$.
\begin{align*}
  \N{\mcd o}& = \{0,1\} \\
  \N{\mcd\alpha}& = \{a | a  {\mbox{ is an $\alpha$-discriminant}}\} \\
  \N{\mcd(\sigma\tau)}& = \mcd\sigma \to \mcd\tau
\end{align*}
We define a value system $\canbe$ as for STT, but extend it to higher types using full function spaces.
\begin{align*}
  \N{s \canbe_o 0}&\iffdef s\in\Wff_o \text{ and } \nf{s}\notin E\\
  \N{s \canbe_o 1}&\iffdef s\in\Wff_o \text{ and } \neg\nf{s}\notin E\\
  \N{s\canbe_\alpha a}\!&\iffdef 
  s\in\Wff_\alpha,~a \text{ is an $\alpha$-discriminant, and } 
  \nf{s}\in a \text{ if } \nf{s} \text{ is discriminating}\\
  \N{\canbe_{\sigma\tau}}&\eqdef\mset{(s,f)\in\Wff_{\sigma\tau}\times(\mcd\sigma\to\mcd\tau)}
  {\forall(t,a)\in\canbe_\sigma\col~(st,fa)\in\canbe_\tau}
\end{align*}
In spite of the slightly different construction, many of the previous results still hold
with essentially the same proofs as before.

\begin{prop}
  \label{efo-prop-norm-poss-value}
  $s\canbe_\sigma a$ iff $\nf{s}\canbe_\sigma a$.
\end{prop}
\begin{proof}  Similar to Proposition~\ref{prop-norm-poss-value}.
\end{proof}

\begin{lem}
  \label{efo-lem-admissibility}
  Let $\mci$ be an assignment into $\mcd$ such that $x \canbe \mci x$
  for all names $x$
  and $\theta$ be a
  substitution such that $\theta{x}\canbe\mci{x}$ for
  all $x\in\Dom\theta$.  Then $s\in\Dom\hat\mci$ and
  $\hat\theta{s}\canbe\hat\mci{s}$ for every term $s$.
\end{lem}
\begin{proof}
  Similar to Lemma~\ref{lem-admissibility}
\end{proof}

\begin{thm}
  \label{theo-efo-admissible-interpretations}
  Let $\mci$ be an assignment into $\mcd$ such that $x \canbe \mci x$
  for all names $x$.
  Then $\mci$ is an
  interpretation such that $s\canbe\hat\mci s$ for all terms $s$.
\end{thm}
\begin{proof}
  Follows from Proposition~\ref{efo-prop-norm-poss-value}, Lemma~\ref{efo-lem-admissibility} and property S4.
\end{proof}

\begin{lem}
  \label{lemma-efo-adm-model}
  A logical assignment $\mci$ is a model of $E$ if $x\canbe \mci x$
  for every name $x$.
\end{lem}
\begin{proof}  Similar to Lemma~\ref{lemma-adm-model}
  using Theorem~\ref{theo-efo-admissible-interpretations}.
\end{proof}

\begin{lem}[Common Value]
  \label{lem-efo-common-value}
  Let $T\incl\Wff_\sigma$.  Then $T$ is compatible if
  and only if there exists a value~$a$ such that
  $T\canbe_\sigma a$.
\end{lem}
\begin{proof}  Similar to Lemma~\ref{lem-common-value}.
\end{proof}

\begin{lem}[Admissibility]
  \label{lemma-efo-inhabitation}
  For every variable $x:\sigma$ there is some $a\in\mcd\sigma$
  such that $x\canbe a$.
\end{lem}
\begin{proof}  Similar to Lemma~\ref{lemma-inhabitation}
  using Lemma~\ref{lem-compatibility} and Lemma~\ref{lem-efo-common-value}.
\end{proof}

\begin{lem}[Functionality]
  \label{lem-efo-functionality}
  If $s\canbe_\alpha a$, $t\canbe_\alpha b$, and
  $(s{=}t)\in E$ , then $a=b$.
\end{lem}
\begin{proof}  Similar to Lemma~\ref{lem-functionality}
  restricted only to sorts.
\end{proof}

As before $\mcl(c)$ is the canonical interpretation for each logical constant $c$.
We now have the additional logical constants $\to$ and $\forall_{\!\alpha}$:
\begin{align*}
  \N{\mcl({\limplies}})&\eqdef\lam{a{\in}\mcd o}{~\lam{b{\in}\mcd o}{~\Cond{a{=}1}b1}}\\
  \N{\mcl({\forall_{\!\alpha}})}&\eqdef\lam{f{\in}\mcd\alpha\to\mcd o}{~\Cond{f = (\lam{x\in\mcd\alpha}{~1})}10}
\end{align*}

\begin{lem}[Logical Constants]
  \label{lem-efo-log-constants}
  $c\canbe\mcl(c)$ for every logical constant $c$. 
\end{lem}
\proof Similar to Lemma~\ref{lem-log-constants}.  The proof for $\neg$ is the same.
  The proof for $\limplies$ uses N3, $\EImp$ and $\EImpN$.
  The proof for $=_\sigma$ requires a slight modification.
  Assume $s\canbe_\sigma a$,
  $t\canbe_\sigma b$, and
  $(s{=_\sigma}t)\ncanbe\mcl(=_\sigma)ab$.  Case analysis.
  \begin{enumerate}[$\bullet$]
  \item $a=b$.  Use Lemmas~\ref{lem-efo-common-value} and~\ref{lem-compatibility}\,(1).
  \item $a\neq b$. Then $(\nf{s}{=}\nf{t})\in E$ and so $\sigma$ must be a sort $\alpha$ since $E$ is EFO.
    This contradicts Lemma~\ref{lem-efo-functionality}.
  \end{enumerate}
  Finally, we prove $\forall_{\!\alpha}\canbe\mcl(\forall_{\!\alpha})$.  Case analysis.  
  Assume $s\canbe_{\alpha o} f$ and $\forall_{\!\alpha} s\ncanbe_o \mcl(\forall_{\!\alpha}) f$.
  \begin{enumerate}[$\bullet$]
  \item $\mcl(\forall_{\!\alpha}) f=1$.  Then $\neg\nf{\forall_{\!\alpha} s} \in E$ and so 
    by N3, $\Ealln$ and N2 we have $\neg\nf{sx}\in E$ for some variable $x:\alpha$.
    We know $\{x\}$ is compatible by Lemma~\ref{lem-compatibility}\,(2) and so by Lemma~\ref{lem-efo-common-value} 
    there is some $a\in\mcd\alpha$ such that $x\canbe a$.
    Thus $sx\canbe 1$, contradicting $\neg\nf{sx}\in E$.
  \item $\mcl(\forall_{\!\alpha}) f=0$.  Then $\nf{\forall_{\!\alpha} s}\in E$ and 
    there is some $a\in\mcd\alpha$ such that $fa = 0$.
    Suppose there are no $\alpha$-discriminating terms.
    In this case $a$ is empty and $u\canbe a$ for any $u\in\Wff_\alpha$.
    By N3, $\Ealld$ and N2 we have $\nf{su}\in E$ for some normal EFO term $u$.
    Hence $su\ncanbe 0$, contradicting $s\canbe f$ and $u\canbe a$.
    Next suppose there are $\alpha$-discriminating terms.
    In this case there is some $u\in a$ by Lemma~\ref{discr-nonempty}.  
    By N3, $\Eall$ and N2 we know $\nf{su}\in E$.
    In this case we also have $su\ncanbe 0$, again contradicting $s\canbe f$ and $u\canbe a$.\qed
  \end{enumerate}

\begin{thm}[EFO Model Existence]
  \label{thm:efo-model-exist}
  Every evident EFO branch has a standard model.
  Every EFO-complete evident EFO branch has a standard model where each $\mcd\alpha$ is countable.
  Every finite evident EFO branch has a finite standard model.
\end{thm}
\begin{proof}
  We use the frame $\mcd$ and relation $\canbe$ defined above.
  We give an assignment $\mci$ into $\mcd$.
  For each variable $x$
  we can choose $\mci x$ such that $x\canbe \mci x$
  using Lemma~\ref{lemma-efo-inhabitation}.
  For each logical constant $c$ we choose $\mci c = \mcl (c)$.
  By Lemma~\ref{lem-efo-log-constants} we know $c\canbe \mci c$.
  $\mci$ is a model of $E$ by Lemma~\ref{lemma-efo-adm-model}.

  Suppose $E$ is EFO-complete.  We prove there are only countably many $\alpha$-discriminants as follows.
  If there are no $\alpha$-discriminating terms, then $\emptyset$ is the only $\alpha$-discriminant.
  Otherwise, every $\alpha$-discriminant is nonempty by Lemma~\ref{discr-nonempty}.  For each $\alpha$-discriminant $a$, choose some $s_a\in a$.
  We prove the function mapping $a$ to $s_a$ is injective.  Assume $a,b\in\mcd\alpha$ and $a\not=b$.
  By EFO-completeness of $E$ and Proposition~\ref{prop-diff-discs} we must have $s_a\not= s_b \in E$.
  If $s_a$ and $s_b$ were the same term, then $E$ would be unsatisfiable.  Hence $s_a$ and $s_b$ are different terms.

  Finally,
  if $E$ is finite, then for each sort $\alpha$ there will be only finitely many $\alpha$-discriminants
  (by Proposition~\ref{prop-finite-discs})
  and hence $\mcd\sigma$ will be finite for all $\sigma$.
\end{proof}

\section{EFO Completeness}\label{sec:efo-complete}

Let $\N{\TSF}$ be the tableau calculus given by taking
all the rules from Figure~\ref{fig:rulesefo} 
subject to the following restrictions.
\begin{enumerate}[$\bullet$]
\item If $(s{\neq} t)$ is on a branch $A$, then
  $\TRFFE$ can only be applied 
  if there is no variable $x$ such that 
  $(\nf{sx}\neq\nf{tx})\in A$.
\item If $\neg\forall_{\!\alpha} s$ is on a branch $A$, then $\TRFalln$ can only be applied if 
  there is no variable $x:\alpha$ such that $\neg\nf{sx}\in A$.
\item If $\forall_{\!\alpha} s$ is on a branch $A$ and there are $\alpha$-discriminating terms in $A$,
  then $\TRFall$ can only be applied with these $\alpha$-discriminating terms. 
\item If $\forall_{\!\alpha} s$ is on a branch $A$,
  $\nf{su}\notin A$ for all normal $u\in\Wff_\alpha$,
  some variable $x:\alpha$ occurs free in $A$
  and there are no $\alpha$-discriminating terms in $A$,
  then $\TRFall$ can only be applied with a variable $x:\alpha$ occurring free in $A$.
\item If $\forall_{\!\alpha} s$ is on a branch $A$,
  $\nf{su}\notin A$ for all normal $u\in\Wff_\alpha$,
  no variable $x:\alpha$ occurs free in $A$
  and there are no $\alpha$-discriminating terms in $A$,
  then $\TRFall$ can only be applied with a variable $x:\alpha$.
\end{enumerate}
The idea behind the restrictions on $\TRFall$ is that only $\alpha$-discriminating terms should
be used as instantiations, except when there are no $\alpha$-discriminating terms.
In case there are no $\alpha$-discriminating terms, at most one new variable $x:\alpha$
will be used as an instantiation term for each sort $\alpha$.
These restrictions will ensure that $\TSF$ terminates when given branches in certain fragments of EFO.

From now on we use the term \emph{refutable} to refer to refutability in the calculus $\TSF$.
That is, the set of \emph{refutable branches} is the least set such that
if $A/\ddd A n$ is
an instance of a rule of~$\TSF$ and $\dd A n$ are
refutable, then $A$ is refutable.
The proof of soundness of $\TS$ (see Proposition~\ref{prop:ts-sound}) extends to show soundness of $\TSF$.

\begin{prop}[Soundness of $\TSF$]
  \label{prop-e-soundness}~
  Every refutable branch is unsatisfiable.
\end{prop}

An EFO abstract consistency class is
a set $\Gamma$ of EFO branches such that every branch $A\in\Gamma$ satisfies the
conditions in Figure~\ref{fig:abs-consistency} and also the following conditions:\\
  \begin{tabular}{c>{\raggedright}p{120mm}}
    \emph{\AImp}&If $s\limplies t$ is in $A$, 
    then $A\cup\set{\neg s}$
    or $A\cup\set{t}$ is in $\Gamma$.
    \tabularnewline
    \emph{\AImpN}&If $\neg(s\limplies t)$ is in $A$, 
    then $A\cup\set{s,\neg t}$ is in $\Gamma$.
    \tabularnewline
    \emph{\Aall}&If $\forall_{\!\alpha} s$ is in $A$,
    then $A\cup\set{\nf{su}}$ is in $\Gamma$ for every $\alpha$-discriminating~$u$ in~$A$.
    \tabularnewline
    \emph{\Aalld}&If $\forall_{\!\alpha} s$ is in $A$,
    then $A\cup\set{\nf{su}}$ is in $\Gamma$ for some normal EFO term $u\in\Wff_\alpha$.
    \tabularnewline
    \emph{\Aalln}&If $\neg \forall_{\!\alpha} s$ is in $A$, 
    then $A\cup\set{\neg\nf{sx}}$ is in $\Gamma$ for some variable $x$.
  \end{tabular}\\[2mm]
We say an abstract consistency class $\Gamma$ is \emph{EFO-complete}
if for all $A\in\Gamma$ and all normal $s\in\EFO_o$ either $A\cup\{s\}\in \Gamma$ or $A\cup\{\neg s\}\in \Gamma$.

Let~\emph{$\GammaTEFO$} be the
set of all finite EFO branches that are not refutable.
\begin{lem}
  \label{lem:acc-efo-completeness}
  $\GammaTEFO$ is an abstract consistency class.
\end{lem}
\proof  Similar to Lemma~\ref{lem:acc-completeness}.
  We only check the new conditions: $\AImp$, $\AImpN$, $\Aall$,
  $\Aalld$ and $\Aalln$.
\begin{enumerate}[\AImpN]
\item[{\AImp}] Let $s\limplies t\in A\in \GammaTEFO$.
  Suppose $A\cup\{\neg s\}\notin\GammaTEFO$ and $A\cup\{t\}\notin\GammaTEFO$.
  By $\TRFImp$ we have $A$ is refutable.  Contradiction.
\item[{\AImpN}] If $\neg(s\to t)\in A$ and $A\cup\{s,\neg t\}\notin\GammaTEFO$,
  then $A\notin\GammaTEFO$ using the rule $\TRFImpN$.
\item[{\Aall}] Let $\forall_{\!\alpha} s\in A\in\GammaTEFO$.
  Suppose $A\cup\{\nf{su}\}\notin\GammaT$ for some normal $\alpha$-discriminating $u$.
  Then $A\cup\{\nf{su}\}$ is refutable.
  Hence $A$ can be refuted using \TRFall (with the restriction).
\item[{\Aalld}] Let $\forall_{\!\alpha} s\in A\in\GammaTEFO$.
  If there is some $\alpha$-discriminating term, then $\Aalld$ follows from $\Aall$.
  Assume there are no $\alpha$-discriminating terms
  and $A\cup\{\nf{su}\}\notin\GammaT$ for all normal $u\in\EFO_\alpha$.
  In particular, $\nf{su}\notin A$ for all normal $u\in\EFO_\alpha$.
  Choose a variable $x:\alpha$ occurring free in $A$ (or any variable $x:\alpha$
  if none occurs free in $A$).
  Since $A\cup\{\nf{sx}\}\notin\GammaT$, $A\cup\{\nf{sx}\}$ is refutable.
  Using $\TRFall$ (with the restriction), $A$ is refutable.  Contradiction. 
\item[{\Aalln}] Let $\neg\forall_{\!\alpha} s\in A\in\GammaTEFO$.  Suppose
  $A\cup\set{\neg\nf{sx}}\notin\GammaT$ for every variable $x:\alpha$.
  Let $x:\alpha$ be fresh for $A$.
  Then $A\cup\set{\neg\nf{sx}}$ is refutable
  and so $A$ can be refuted using $\TRFalln$.\qed
\end{enumerate}

\begin{lem}[EFO Extension Lemma]
  \label{lem:efo-extension}
  Let $\Gamma$ be an abstract consistency class and
  $A\in\Gamma$ be an EFO branch.  Then there exists an evident EFO branch
  $E$ such that $A\subseteq E$.  Moreover, if $\Gamma$
  is EFO-complete, a EFO-complete evident EFO branch $E$ exists
  such that $A\subseteq E$.
\end{lem}
\begin{proof}  Similar to Lemma~\ref{lem:extension}.  Instead of using an enumeration of all normal formulas,
  we use an enumeration of all normal EFO formulas.  The proof goes through when one makes
  some obvious modifications.
\end{proof}

\begin{thm}[EFO Completeness]
  \label{thm:efo-completeness}
  Every finite EFO branch is either refutable
  or has a standard model.
\end{thm}
\begin{proof}
  Follows from Lemma~\ref{lem:acc-efo-completeness}, Lemma~\ref{lem:efo-extension} and
  Theorem~\ref{thm:efo-model-exist}.
\end{proof}

We now turn to compactness and the existence of countable models.
Let $\GammacompEFO$ be the set of all sufficiently pure
EFO branches $A$ such that every finite subset of $A$ has a standard model.

\begin{lem}
  \label{lem:acc-efo-compactness}
  $\GammacompEFO$ is an EFO-complete abstract consistency
  class.
\end{lem}
\begin{proof}  Similar to Lemma~\ref{lem:acc-compactness}.
\end{proof}

\begin{thm}\label{thm:efo-compactness-ls}
  Let $A$ be a branch such that every finite subset of
  $A$ has a standard model.  Then $A$ has a standard model where $\mcd\alpha$ is countable for all sorts $\alpha$.
\end{thm}
\begin{proof}  Similar to Theorem~\ref{thm:compactness-ls}.
\end{proof}

\begin{cor}\label{cor:efo-stdsat}
  Let $A$ be a satisfiable EFO branch.
  Then $A$ has a standard model where $\mcd\alpha$ is countable for all sorts $\alpha$.
\end{cor}
\begin{proof}
  To apply Theorem~\ref{thm:efo-compactness-ls}
  we only need to show every subset of $A$ has a standard model.
  Let $A'$ be a finite subset of $A$.
  Since $A'$ is satisfiable, $A'$ is not refutable by Proposition~\ref{prop-e-soundness}.
  By Theorem~\ref{thm:efo-completeness} $A'$ has a standard model.
\end{proof}

\section{Decidable EFO Fragments}

Given the completeness result for the tableau calculus $\TSF$
(Theorem~\ref{thm:efo-completeness}),
we can show a fragment of EFO is decidable by proving $\TSF$
terminates on branches in the fragment.  We will use this technique to argue
decidability of three fragments:
\begin{enumerate}[$\bullet$]
\item The \emph{$\lambda$-free fragment}, which is EFO without $\lambda$-abstraction.
\item The \emph{pure fragment}, which consists of disequations $s\neq t$ where no name used in $s$ and $t$ has a type that contains $o$.
\item The \emph{BSR fragment (Bernays-Sch\"onfinkel-Ramsey)}, which consists of relational first-order $\exists^*\forall^*$-formulas with equality.
\end{enumerate}

\begin{prop}[Verification Soundness]
  \label{prop-verif-soundness}
  Let $A$ be a finite EFO branch that is not closed and
  cannot be extended with $\TSF$.  Then $A$
  is evident and has a finite model.
\end{prop}
\begin{proof}
  Checking $A$ is evident is easy.  The existence of a finite model follows from Theorem~\ref{thm:efo-model-exist}.
\end{proof}

We now have a general method for proving decidability of satisfiability
within a fragment.

\begin{prop}
  Let $\TSF$ terminate on a set $\Delta$ of finite
  EFO branches.  Then satisfiability of the branches in
  $\Delta$ is decidable and every satisfiable branch in
  $\Delta$ has a finite model.
\end{prop}

\begin{proof}
  Follows with Propositions~\ref{prop-e-soundness}
  and~\ref{prop-verif-soundness} and
  Theorem~\ref{thm:efo-model-exist}.
\end{proof}

The decision procedure depends on the normalization
operator employed with $\TSF$.  A normalization
operator that yields $\beta$-normal forms provides for
all termination results proven in this section.  Note
that the tableau calculus applies the normalization
operator only to applications $st$ where $s$ and $t$
are both normal and $t$ has type $\alpha$ (for some sort $\alpha$) if it is not a
variable.  Hence at most one $\beta$-reduction is
needed for normalization if $s$ and $t$ are
$\beta$-normal.  Moreover, no $\alpha$-renaming is
needed if the bound variables are chosen differently
from the free variables.  For clarity, we continue to
work with an abstract normalization operator and state
further conditions as they are needed.

\subsection{Lambda-Free Formulas}

In~\cite{BrownSmolkaBasic} we study lambda- and
quantifier-free EFO and show that the concomitant
subsystem of $\TSF$ terminates on finite branches.  The
result extends to lambda-free branches containing
quantifiers (e.g., $\set{\forall_{\!\alpha} f}$).

\begin{prop}[Lambda-Free Termination]
  Let the normalization operator satisfy\lmcs{\linebreak}
  $\nf{s}=s$ for
  every lambda-free EFO term $s$.  Then $\TSF$
  terminates on finite lambda-free branches.
\end{prop}

\begin{proof}
  An application of \TRFFE disables a disequation
  $s{\neq_{\sigma\tau}}t$ and introduces new subterms
  as follows: a variable $x:\sigma$, two terms
  $sx:\tau$ and $tx:\tau$, and the formula 
  $sx{\neq}tx$.  The types of the new
  subterms are smaller than the type of $s$ and $t$,
  and the new subterms introduced by the other rules
  always have type $o$ or $\alpha$.
  For each branch,
  consider the multiset of types
  $\sigma\tau$ where
  $s,t:\sigma\tau$ are subterms of formulas on the branch but
  there is no $x:\sigma$ such that
  $sx\neq tx$ is on the branch.
  By considering the multiset ordering, we see that no derivation can
  employ \TRFFE infinitely often.

  Let $A\to A_1\to A_2\to\cdots$ be a possibly infinite
  derivation that issues from a finite lambda-free
  branch and does not employ \TRFFE.  It suffices to
  show that the derivation is finite.  
  Consider the new variables $x:\alpha$ which may be introduced
  by the $\TRFall$ and $\TRFalln$ rules.
  For each subterm $\forall_{\!\alpha} s$ at most one new variable will
  be introduced by these rules.
  Since the branches are $\lambda$-free, no rule creates new subterms of the form $\forall_{\!\alpha} s$.
  Hence only finitely many new variables of type $\alpha$
  are introduced.
  Let $A_n$ be a branch in the sequence such that no new
  variables are introduced after this point.  
  Let $S_\sigma$ be the set of all subterms of type $\sigma$ of the formulas in $A_n$.
  Let $B$ be the union of the three finite sets 
  $S_o$, $\{\neg s | s\in S_o\}$
  and $\{s\not=_\sigma t | s,t\in S_\sigma\}$.
  Every branch $A_m$ with $m\geq n$
  can only contain members of $B$.  Hence the derivation is finite.
\end{proof}

\subsection{Pure Disequations}

A type is \emph{pure} if it does not contain $o$.  A
term is \emph{pure} if the type of every name occurring
in it (bound or unbound) is pure.  An equation $s=t$ or
disequation $s\neq t$ is \emph{pure} if $s$ and $t$ are
pure terms.

We add a new property of normalization in order to prove termination.
\begin{description}
\item[{N5}] 
  The least relation $\succ$ on terms such that
  \begin{enumerate}[(1)]
  \item ${a\ddd s n}\succ{s_i}$ \ if 
    $i\in\set{1\cld n}$
  \item $s\succ\nf{sx}$ \ if $s:\sigma\tau$ and $x:\sigma$
  \end{enumerate}
  terminates on normal terms.
\end{description}

\begin{prop}[Pure Termination]
  Let the normalization operator satisfy N5.  Then
  $\TSF$ terminates on finite branches containing only
  pure disequations.
\end{prop}

\begin{proof}
  Let $A\to A_1\to A_2\to\cdots$ be a possibly infinite
  derivation that issues from a finite branch
  containing only pure disequations.  Then no other
  rules but possibly \TRFDec and \TRFFE apply
  and thus no $A_i$ contains a formula that is not
  a pure disequation (using S5).  Using N5
  it follows that the derivation is finite.
\end{proof}

\subsection{Bernays-Sch\"onfinkel-Ramsey Formulas}

It is well-known that the satisfiability of Bernays-Sch\"onfinkel-Ramsey
formulas (relational first-order $\exists^*\forall^*$-prenex
formulas with equality) is decidable and the fragment has the
finite model property~\cite{BGG97}.  We reobtain this
result by showing that $\TSF$ terminates for the
respective fragment.  We call a type \emph{BSR} if it
is $\alpha$ or $o$ or has the form $\alpha_1\dots\alpha_n o$.
We call an EFO formula $s$ \emph{BSR} if it satisfies
two conditions:
\begin{enumerate}
\item The type of every variable that occurs in $s$ is
  BSR.
\item $\forall_{\!\alpha}$ does not occur below a negation or an implication in
  $s$.
\end{enumerate}
Note that every subterm of a BSR formula that has type $\alpha$
is a variable.
For simplicity, our BSR formulas don't provide for
outer existential quantification.  We need one more
condition for the normalization operator:
\begin{description}
\item[{N6}] If $s:\alpha o$ is BSR and $x:\alpha$,
  then $\nf{sx}$ is BSR.
\end{description}

\begin{prop}[BSR Termination]
  Let the normalization operator satisfy N5 and~N6.
  Then $\TSF$ terminates on finite branches containing
  only BSR formulas.
\end{prop}

\begin{proof}
  Let $A\to A_1\to A_2\to\cdots$ be a possibly infinite
  derivation that issues from a finite branch
  containing only BSR formulas.  Then \TRFalln and
  \TRFFE are not applicable and all $A_i$ contain only
  BSR formulas (using N6).  Furthermore, for each sort $\alpha$
  used in $A$ at most one new
  variable of sort $\alpha$ is introduced (by the restriction on $\TRFall$ in $\TSF$).
  Since all terms of sort 
  $\alpha$ are variables, there is only a finite supply.
  Using N5 it follows that the derivation is finite.
\end{proof}

\section{Conclusion}

In this paper we have studied a complete cut-free tableau calculus
for simple type theory with primitive equality (STT).  For the first-order
fragment of STT (EFO) we have shown that the tableau system is complete with respect
to standard models.  Our development demonstrates
that first-order logic can be treated naturally as a fragment of STT.

For the EFO fragment we gave an interesting restriction on instantiations.
In particular, one can restrict most instantiations of sort $\alpha$ to be $\alpha$-discriminating terms.
Such a restriction can also be included in the tableau calculus for STT without sacrificing
completeness.  Confining instantiations to $\alpha$-discriminating terms
is a serious restriction since each branch has only finitely many such terms.

Automated theorem proving would be a natural application of the tableau calculi presented here.
When designing a search procedure one often starts with a complete ground calculus (like our
tableau calculi $\TS$ and $\TSF$), then extends this to include metavariables to be instantiated
during search, and finally proves a lifting lemma showing the tableaux with metavariables can
simulate a refutation in the ground calculus.
A design principle of our calculi $\TS$ and $\TSF$ is that none of the
rules look deeply into the structure of any formula on the branch.
For example, consider the mating rule
\begin{mathpar}
  \inferrule*[right=~$n\geq 0$] {xs_1\dots
    s_n\,,\,\neg xt_1\dots t_n} {s_1\neq t_1\mid\dots\mid s_n\neq t_n}
\end{mathpar}
To check if this rule applies to two formulas $s,t$ on the branch $A$,
one only needs to check if $s$ has a variable $x$ at the head
and if $t$ is the negation of a formula with $x$ at the head.
When trying to prove a lifting lemma, we would need to show how the 
calculus with metavariables could simulate the mating rule.
This may involve partially instantiating metavariables to expose
the head $x$ in the counterpart to $s$ or
the negation and the head $x$ in the counterpart to $t$.
On the other hand, suppose our ground calculus included a rule to close branches with 
a formula of the form $s\not= s$.
To simulate this in the calculus with metavariables we would need to know if
some instantiation for the metavariables can yield a formula of the form $s\not=s$.
In the worst case this is a problem requiring full higher-order unification.
We have been careful to only include rules in our calculi which will
not require arbitrary instantiations of metavariables to prove a lifting lemma.
Formulating such a calculus with metavariables and proving such a lifting lemma
is left for future work.



\bibliographystyle{plain}

\end{document}
